\pdfoutput=1
\documentclass[twocolumn]{aastex62}%
\usepackage{amsmath}
\usepackage{mathtools}
\usepackage{multirow}
\usepackage{booktabs}
\usepackage{bbding}
\usepackage{bm}
\usepackage{amssymb}

\begin{document}


\title{UNVEILING NON-GRAY SURFACE OF CLOUDY EXOPLANETS: THE INFLUENCE OF WAVELENGTH-DEPENDENT SURFACE ALBEDO AND CLOUD SCATTERING PROPERTIES ON RETRIEVAL SOLUTIONS}


\correspondingauthor{Jinping He}
\email{jphe@niaot.ac.cn}

\author{Fei Wang}
\affiliation{National Astronomical Observatories/Nanjing Institute of Astronomical Optics $\&$ Technology, Chinese Academy of Sciences, Nanjing, 210042,China.}\affiliation{CAS Key Laboratory of Astronomical Optics $\&$ Technology, Nanjing Institute of Astronomical Optics $\&$ Technology, Nanjing, 210042, China. }\affiliation{University of Chinese Academy of Sciences, Beijing, 100049,China.}

\author{Yuka Fujii}
\affiliation{Division of Science, National Astronomical Observatory of Japan, Osawa, Mitaka, Tokyo 181-8588, Japan.}\affiliation{Earth-Life Science Institute, Tokyo Institute of Technology, Ookayama, Meguro, Tokyo 152-8550, Japan.}

\author{Jinping He}
\affiliation{National Astronomical Observatories/Nanjing Institute of Astronomical Optics $\&$ Technology, Chinese Academy of Sciences, Nanjing, 210042,China.}\affiliation{CAS Key Laboratory of Astronomical Optics $\&$ Technology, Nanjing Institute of Astronomical Optics $\&$ Technology, Nanjing, 210042, China. }

\begin{abstract}
Direct-imaging spectra hold rich information about a planet's atmosphere and surface, and several space-based missions aiming at such observations will become a reality in the near future. Previous spectral retrieval works have resulted in key atmospheric constraints under the assumption of a gray surface, but the effect of wavelength-dependent surface albedo on retrieval has not been shown. We explore the influence of the coupling effect of cloud and wavelength-dependent surface albedo on retrieval performance via modeling suites of Earth-like atmospheres with varying cloud and surface albedo parameterizations. Under the assumption of known cloud scattering properties, the surface spectral albedos can be reasonably recovered when the surface cover represents that of Earth-like vegetation or ocean, which may aid in characterizing the planet's habitability. When the cloud scattering properties cannot be assumed, we show that the degeneracy between the cloud properties and wavelength-dependent surface albedo leads to biased results of atmospheric and cloud properties. The multi-epoch visible band observations offer limited improvement in disentangling this degeneracy. However, the constraints on atmospheric properties from the combination of UV band (R $\sim 6$) $+$ visible band (R $\sim 140$) are consistent with input values to within 1 $\sigma$. If short bandpass data is not available, an alternative solution to reduce the retrieval uncertainties would be to have the prior constraints on planetary cloud fraction with less than 20\% uncertainty.
\end{abstract}



\received{01-Nov-2021}
\revised{24-Mar-2022}
\accepted{14-Apr-2022}
\submitjournal{The Astrophysical Journal}
\section{Introduction}

Direct imaging offers the possibility of characterizing planets around nearby stars and at relatively large orbital distances, enabling characterization in complementary phase space to the types of planets that transmission spectroscopy can probe. These spectroscopy observations are encoded with rich information on the chemical composition and clouds in exoplanet's atmospheres. Unlike transmission spectroscopy, direct-imaging spectroscopy in the visible band could probe atmospheric depths of up to $\sim$ 10 bar or even the surface of terrestrial exoplanets. The interpretation of these spectra will revolutionize planetary astronomy and spur decades of work in observational astrobiology. 

The development of the technology that will enable these observations is rapidly progressing. Next-generation ground-based telescopes such as the TMT, ELT, etc., will be capable of performing high-contrast imaging and high-resolution spectroscopy on exoplanets from gas giants down to sub-Neptunes \citep{snellen2015,quanz2015,mazin2019}. Upcoming space-based direct-imaging mission concepts (e.g., Habitable Exoplanet Observatory [HabEx], the Large UV/Optical/IR Surveyor [LUVOIR]) could reach a contrast ratio of order $10^{-10}$, and would perform low-resolution spectroscopy of terrestrial exoplanets orbiting nearby stars \citep{fujii2018,dressing2019,roberge2019,roberge2021,checlair2021}. Although the above space and ground-based telescopes are still years away, it is crucial to investigate what information can be extracted from measurements of reflected starlight from spatially unresolved planets. Such investigations will have an impact on the design of future telescopes and their observing strategies.

Direct-imaging observations of exoplanets in reflected starlight are interpreted mainly via the Bayesian retrieval technique to constrain various atmospheric properties from an observed spectrum. Major advancements in atmospheric retrieval have been made in the last decade, which resulted in key constraints on the atmospheric properties for several exoplanets (e.g., \citealp{cahoy2010,greco2015,lupu2016,robinson2016,nayak2017,macdonald2017hd,batalha2018,feng2018,guimond2018,lacy2019,damiano2020,damiano2020multi,mukherjee2021,carrion2020,carrion2021}). Specifically, the molecular absorptions, the cloud properties (e.g., its thickness, the pressure level of the cloud top, the scattering phase function, the single-scattering albedo) and the surface albedo are the main factors that affect a reflected-light spectrum \citep{stephens2000,carrion2020}. There have been studies that explored the retrieval prospect with future direct imaging mission to place constraints on basic cloud properties for cool giants under the assumption of a dark surface \citep{lupu2016} and characterized Earth analogs in reflected light with assumed optical properties of water cloud and a gray surface with the HabEx/LUVOIR-type instruments \citep{feng2018}. All of these studies offer insights into the kind of constraints we expect on the exoplanets' properties from high-contrast imaging-era spectra. However, in future exoplanet spectroscopy retrieval, not only the optical properties of clouds but also the wavelength dependence of the surface reflectivity is likely to be unknown. Previous spectral retrieval work characterized the planetary surface with a gray (i.e., wavelength-independent) surface albedo, which may result in retrieval bias for exoplanets with non-gray surface albedo. For example, with the presence of oceans and continents, the Earth's albedo depends on wavelength. In particular, the albedo spectra of vegetation commonly have a ``red edge," which corresponds to a jump in reflectivity around 0.7$\mu$ m \citep{seager2005}. These unique albedo spectra may bring additional information in spectral retrieval. However, they could also be degenerate with other planetary properties (e.g., cloud properties) and bring uncertainty to parameter estimations. 

To mimic the most realistic exoplanets retrieval schemes, we may have to explore the coupling effect of cloud optical properties and the wavelength-dependent surface albedo on retrieval performance and disentangle the possible degeneracy sources in the future direct-imaging spectroscopy. This will help anticipate the challenges that will arise when interpreting future direct-imaging spectra and provide mitigation strategies.

In this paper, we intend to develop a methodology to retrieve cloud properties and the wavelength-dependent surface albedo simultaneously. The mock observations data of low spectral resolution spectroscopy expected from LUVOIR are utilized to test the methodology. In doing this, we specifically aim to address the following:

1. How does the wavelength dependence of albedo disturb our spectral retrieval?

2. How do the unknown cloud scattering properties influence our spectral retrieval?

3. Can the multi-epoch observations or added Rayleigh-bandpass help overcome the difficulty with unknown cloud scattering properties?

4. Which parameters are critical to know or assume a priori for the trustworthy parameter estimation?

Our paper is organized as follows: Section 2 describes the model setup, including the model assumptions (Section 2.1-2.3), the retrieval framework, and the experimental setup (Section 2.4). Section 3 presents the results from the retrieval experiments exploring the effect of wavelength-dependent surface albedo (Section 3.1), unknown cloud scattering properties (Section 3.2), mitigation strategies (Section 3.3), and constraints on surface composition (Section 3.4). We discuss the outstanding degeneracies among cloud and surface parameters (Section 4.1), the requisite prior information to give a trustworthy retrieval for different model parameters (Section 4.2), the model limitations (Section 4.3), and finally summarize the results in Section 5.
\section{Model Setup} 

\subsection{Assumptions and parameterizations in calculating reflectance spectra}
We use PICASO \citep{batalha2019}, which heritages from classic atmospheric radiative transfer code \citep{mckay1989,marley1999}, to model the reflected light spectra of our target planets. PICASO is a well-parameterized radiative transfer code capable of calculating transmission, reflected, and thermal spectra of planets and brown dwarfs. In addition to basic planetary properties (e.g., stellar spectrum, planet mass, and radius), PICASO requires the pressure-temperature (P-T) profile, cloud structure, atmospheric composition, and surface albedo as inputs for the radiative transfer calculation. PICASO uses the two-stream quadrature to solve for the diffuse scattered radiation. For the single-scattering component, in order to incorporate both Rayleigh and the cloud scattering properties, PICASO uses the TTHG$\_$ray as the default phase function. The TTHG$\_$ray weights the TTHG (the two-term Henyey-Greenstein phase function) and Rayleigh scattering (ray) by the fractional opacity of each. The multiple-scattering component of the source function must be integrated over all diffuse angles. In PICASO, an  N = 2 expansions Legendre polynomials are used to compute the integrated source function.
\subsubsection{P-T profile and cloud structure}
Here, we divide the planet's atmosphere into 50 plane-parallel pressure layers where the pressure rises logarithmically from $10^{-4}$ (pressure at the top of the atmosphere, $P_{0}$) to 1 bar (surface pressure, $P_{s}$). For simplicity, we assume an isothermal atmosphere (at T = 250 K), as the temperature has little effect on the low-resolution reflected-light spectrum \citep{robinson2017}.

In \citet{lupu2016}, they characterize surface as a dark surface, and their two-layer cloud model atmosphere includes a deep, optically thick cloud deck that essentially acts as a reflective surface. Unlike the gas giants considered in that study, terrestrial planets have a solid surface we can probe. Thus, we adopt wavelength-dependent surface albedo. For cloudy areas, we assume a single-layer H$_2$O cloud. The horizontal distribution of clouds was characterized by a parameter of $f_{\rm cld}$, the fractional coverage of clouds averaged over the illuminating surface. The geometric albedo spectrum is synthesized by weighted-averaging the ``cloudy" and ``cloud-free" radiative flux. The quantities that define the cloud deck are $p_b$, the cloud-base pressure; $dp$, the pressure difference between the cloud base and cloud top; $\tau$, optical depth; $g_{0}$, scattering asymmetry; and $\omega$, single scattering albedo. 

\begin{figure}[!htb]
	\centering
	\includegraphics[height=12cm,width=9cm]{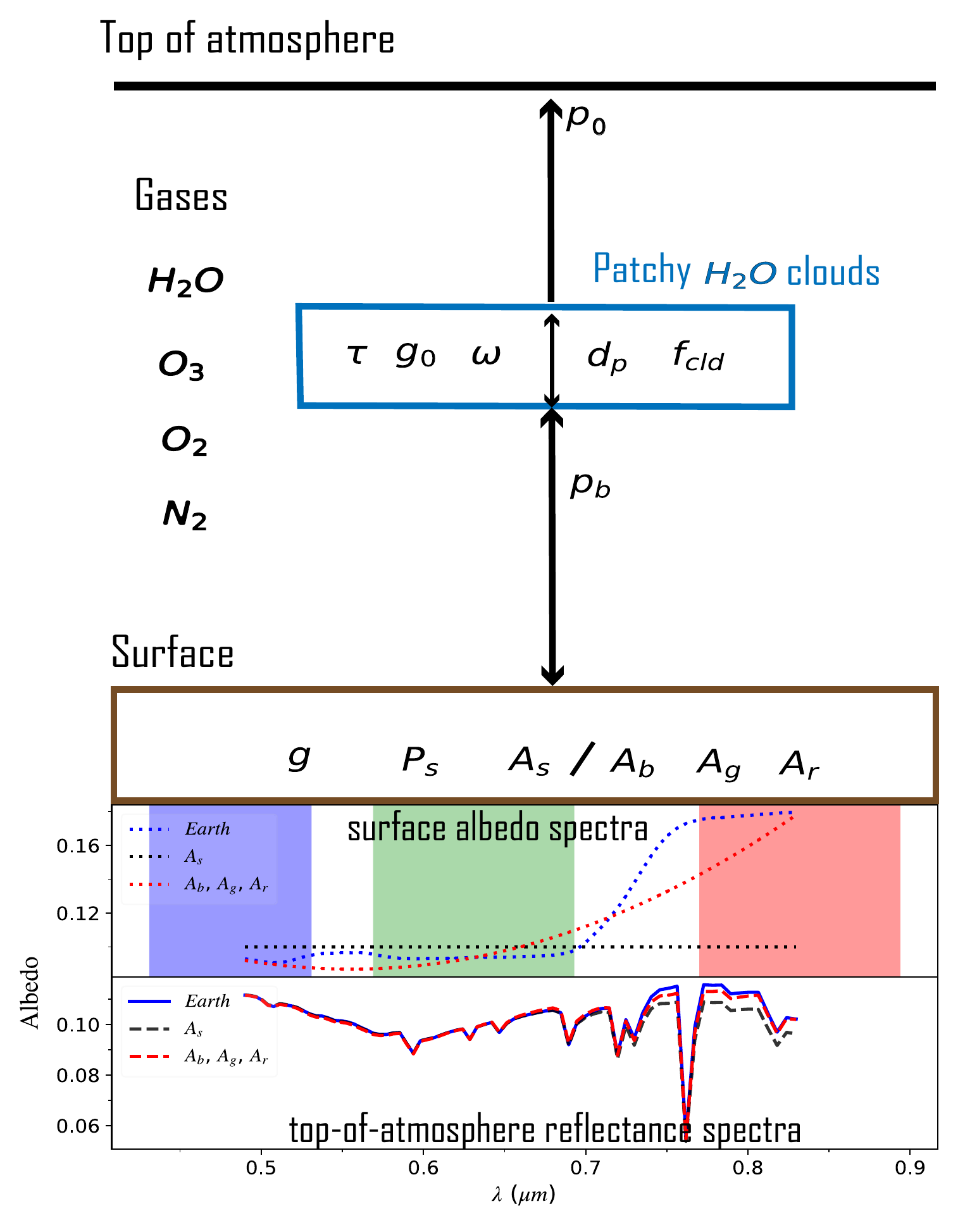}
	\caption{The upper panel shows the illustrative schematic of the atmosphere, cloud structure, and surface parameterization schemes. The middle panel shows the comparison of the gray surface albedo scheme ($A_s$) to our three photometric bands albedo scheme ($A_{b}$, $A_{g}$, $A_{r}$). The black dotted lines stand for the $A_s$. The colored shadow region represents the three photometric bands used to characterize surface albedo. The interpolated surface albedo spectrum (red dotted curve) is compared to the reference Earth surface albedo spectrum (blue dotted curve, R = 140). The bottom panel shows the simulated reflectance spectra using the $A_s$ (black dashed curve) and the 3-bands albedo (red dashed curve). The reflectance spectrum generated with the Earth's surface albedo is given as a reference (blue curve). The reflectance spectra generated with the two surface parameterization schemes are similar while showing deviation in the red band compared with the reference spectrum. Please refer to Section 2.1.3 in the text for a detailed description.}
	\label{img1}%
\end{figure} 

\subsubsection{Atmospheric composition}

We focus on molecular absorption due to H$_2$O, O$_2$, and O$_3$. The primary Rayleigh scatterer and background gas in our fiducial model is N$_2$, whose abundance makes up the remainder of the atmosphere after all other gases are accounted for.

\subsubsection{Surface albedo parameterizations}

Previous retrieval work characterized the planetary surface with a gray surface albedo ($A_{s}$), which represents the specific power in scattered, outgoing radiation compared to that in incident radiation \citep{lupu2016,feng2018}. However, terrestrial planets, in general, have wavelength-dependent albedo given a wide range of different exoplanets surface covers \citep{ hu2012,fujii2014}. For instance, the surface albedo has the wavelength response on an Earth-like terrestrial exoplanet featuring oceans and continents. In particular, as shown in Figure 1 (the second panel from the bottom), Earth's surface albedo spectrum has an uplift at longer wavelengths. This shape is partly biogenic because vegetation's red edge makes the spectrum red, while the ocean absorbs visible light, creating the low reflectance bottom at short wavelengths.

Here, we adopted three optimized photometric bands for identifying Earth-like exoplanets to represent the wavelength-dependent albedo. The three optimal photometric bins are inherited from \citet{krissansen2016} for effectively separating Earth from uninhabitable planets. The surface albedo of the three photometric bands is parameterized as $A_{b}$ (431–531 nm), $A_{g}$ (569–693 nm),  and $A_{r}$ (770–894 nm). We note that these parameters represent the albedo at the central wavelength of each photometric band (shadowed areas in Figure 1), and then are cubic spline interpolated into a given resolution spectrum (R = 140 in this paper, red dotted curve) to reflect the wavelength-dependent surface albedo. The reference surface albedo spectrum  (blue dotted curve) is computed via weighted-averaging the reflectance response of different surface types on Earth. As shown in Figure 1 (the second panel from the bottom), the interpolated surface albedo spectrum could capture the overall albedo variation with wavelength. The corresponding reflectance spectrum (red dashed curve in the bottom panel of Figure 1) generated with the interpolated surface albedo is consistent with the reference spectrum (blue curve). In contrast, the gray surface albedo scheme ($A_{s} = 0.1$) failed to fit the reference spectrum at the red band since its wavelength-independent low albedo response.

The illustrative parameterization schemes of the atmosphere, cloud structure, and the surface albedo are summarized in Figure 1.

\subsection{Forward model sensitivity tests}

Before performing any retrievals, we first attempt to build our intuition for how the various parameters of the forward model influence the observed spectrum. Tests of this manner could bring to light any potential biases or degeneracies in the results that could impact the interpretation of real spectra. Figure 2 and items (a–f) summarize these effects.

\begin{figure*}[!htb]
	\centering
	\includegraphics[height=12cm,width=18cm]{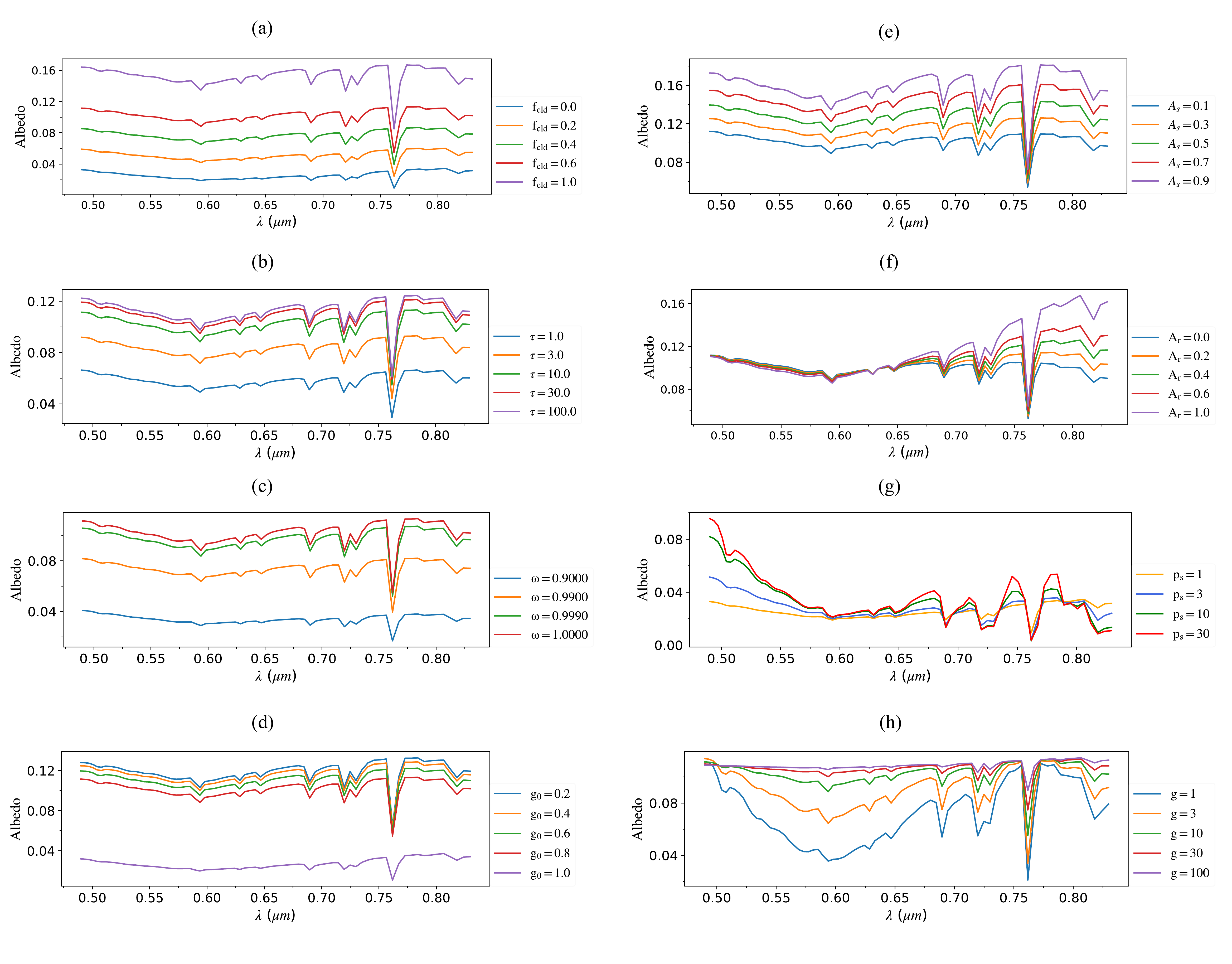}
	\caption{Parameter sensitivity tests. The colored lines show the effects of changing the eight parameters in our forward model: (a) $f_{\rm cld}$, (b) $\tau$, (c) $\omega$, (d) $g_{0}$, (e) $A_{s}$, (f) $A_{r}$, (g) $p_{s}$, and (h) g. The fiducial values used in these simulations are given in Table 1, except for the panel (g), whose atmosphere is cloudless ($f_{\rm cld}$ = 0).}
	\label{img2}
\end{figure*}

(a) $f_{\rm cld}$: Cloud cover fraction. $f_{\rm cld}$ is the cloudiness of the observed planetary disk. $f_{\rm cld}$ directly changes the overall observed reflectance. It can also change the depth of the absorption bands by scattering light back to space before it reaches the absorber and/or by increasing the average optical path through the atmosphere \citep{fauchez2017,fujii2013}. 

(b) $\tau$: Cloud optical thickness. The optical thickness is the degree to which the cloud prevents light from passing through it. The higher a cloud's optical thickness, the more sunlight the cloud is scattering and reflecting, which leads to higher spectra albedo.

(c) $\omega$: Single scattering albedo of clouds (value between 0 - 1). The single scattering albedo of cloud affects the overall albedo of the reflectance spectrum. As shown in panel c of Figure 2, the characteristic $\omega = 1$ of water cloud discriminates itself clearly from other values in their reflected light spectra. 

(d) $g_{0}$: Asymmetry factor of clouds used in TTHG\_ray phase function in PICASO. The asymmetry parameter $g_{0}$ ranges between -1 and 1. In the limit of $g_{0}$ = 1, photons approximately continue traveling in their original direction; when $g_{0}$ = -1, their directions are reversed; and when $g_{0}$ = 0, they are equally likely to travel in the forward or backward directions. Generally, larger $g_{0}$ will lead to lower spectra albedo. 

(e-f) $A_s$ and $A_{r}$: Gray surafce albedo and surface albedo in the red band (770–894 nm). The surface reflectivity is one of the main parameters that affect reflected lights. The gray surface albedo of $A_{s}$ controls the overall reflectivity across the whole bandpass. Thus, it is easily degenerate with the cloud properties of $f_{\rm cld}$, $\tau$, $\omega$, and $g_{0}$. $A_{r}$ represents our 3-bands surface albedo parameterization. The 3-bands surface albedo parameterization reflects the albedo variation between the three photometric bands. Their effect on reflectance spectra is wavelength-dependent. The reflectance spectrum of the Earth is upward sloping in $A_{r}$ bands due to its relatively high reflectivity of continents, which includes some vegetation components (\citealt{arnold2002,seager2005}).

(g-h) $g$ and $P_{s}$: Surface gravity and surface pressure. Both gravity and surface pressure influence the column mass, which controls, e.g., the appearance of spectral features due to gaseous absorption and the Rayleigh-scattering. Besides, $P_{s}$ also affects the line shape by the effect of pressure broadening. 

\subsection{Fiducial values}

The well-parameterized albedo model PICASO enables the computation of a wide diversity of exoplanet and brown dwarf spectra in transmission, emission, or reflected light. Here we define a set of fiducial input parameters designed to mimic Earth-like exoplanets with non-gray surface cover and cloudy atmosphere, enabling us to generate simulated observational data sets for retrieval.

Table 1 summarizes the fiducial model parameter values adopted for our Earth twin. Also shown are the priors for these parameters, which are used to perform retrieval analysis. Briefly, these parameters can be divided into four categories: planetary bulk properties, atmospheric composition, cloud properties, and surface properties. The surface pressure is set at $P_{s}$ = 1 bar and we adopt a surface gravity of g = 9.8 $ \rm m\ s^{-2}$, which is representative of Earth-twin. The representative molecular absorption due to H$_2$O, O$_2$, and O$_3$ are included. The input mixing ratios values are H$_2$O = $3\times 10^{-3}$, O$_3$ = $7 \times 10^{-7}$, and O$_2$ = 0.21.

We allow for a single cloud deck with homogeneous distribution but varying cloud cover fractions in the atmosphere. The vertical distribution of cloud is bounded by cloud base pressure $P_{b}$ = 0.6 bar and $p_{top}$=$P_{b}$+$d_{p}$ ($d_{p}$ = 0.1 bar). Cloud optical depth ($\tau$) and cloud fraction ($f_{\rm cld}$) were set to 10 and 0.6, respectively, based on the MODIS data \citep{king1997}. The water cloud on Earth has a characteristic scattering asymmetry $g_{0} \approx$ 0.8, and single scattering albedo $\omega$ = 1, which are determined by cloud particle effective radius. We fixed these two values of $\omega$ and $g_{0}$ in our retrieval experimental model I and then took them as fitting parameters to explore their effect on retrieval in experimental models II-V.

In computing the model spectrum to retrieve, we used the medium resolution spectra of surface albedo (blue curve of the middle panel of Figure 1). This gives the following values for the 3-bands albedo: $A_{b}$ = 0.09, $A_{g}$ = 0.09, $A_{r}$ = 0.18. In the retrieval model that adopts 3-bands albedo parameterization, we use these values as the fiducial values. 

\subsection{Noise model and retrieval setup}

As a proxy of the observed data, we created mock spectra taking account of the expected observational noises. The mock observation is performed under the next-generation direct-imaging spectroscopic instrument assembled in a LUVOIR-A-like telescope. The designed spectral resolution of the visible band is assumed to be 140. We compute the model spectrum with the continuous medium resolution surface albedo spectrum (R = 1000) and then degrade this medium-resolution albedo spectrum to the resolution of 140. We added the observational noises to the spectrum under the assumption of LUVOIR concepts observation given the signal-to-noise ratio (S/N).

Notably, to mitigate the influence of random noise instances, we did not randomize the displacement of data points for any simulated spectra. In other words, the individual simulated spectral points are placed on the ``true” reflectance albedo point and are assigned error bars according to the S/N. This is because a single random noise instance in the data can introduce unpredictable biases in the posteriors of retrieved parameters. In principle, we need to combine and average posterior distributions of retrievals obtained with a large number ($\ge$ 10) of simulated data sets at a given model setup to eliminate biases stemming from Gaussian noises. However, the computational limitations prevent us from doing this \citep{caldas2019}. As was pointed out in \citet{feng2018}, the average posteriors of retrievals from multiple random noise instances are close to the ones from nonrandomized data, which can be indicated from the central limit theorem.

\begin{table*}[!htb]
	\centering
	\caption {The Retrieved Parameters and Prior Ranges for the $C_{\rm \omega, g_{0}}S_{\rm 3bands}$ Model}
	\begin{tabular}{@{}lllll@{}}
		\toprule
		\toprule
		Parameter variant & Parameter & Description  & Input Value & Prior ranges \\ \midrule
		\multirow{2}{*}{Bulk properties}                                                         &  log$P_{s}$ ($\rm bar$)                      & Surface pressure                & log(1)                            & {[}-2,2{]}                       \\
		& log$g$ ($\rm m\ s^{-2}$)                          & Surface gravity           & log(9.8)                             & {[}0,2{]}   \\ \midrule                      
		\multirow{3}{*}{Atmospheric properties}  &    logH$_2$O                          & Water vapor mixing ratio        & log(3e-3)                            & {[}-8,1{]}                       \\
		& logO$_2$                           & Oxygen mixing ratio             & log(0.21)                            & {[}-10,0{]}                      \\
		& logO$_3$                           & Ozone mixing ratio              & log(7e-7)                            & {[}-10,-1{]}  \\ \midrule                    
		\multirow{6}{*}{Cloud Properties}                                                        & log$P_{b}$                          & Cloud base pressure             & -0.22                           & {[}-2,2{]}                       \\
		& log$d_{p}$                            & Cloud thickness                 & -1                              & {[}-2,2{]}                       \\
		& log$\tau$                          & Cloud optical thickness         & log(10)                              & {[}-1,2{]}                       \\
		
		& log$\omega$                        & Cloud single scattering albedo &
		log(1)                               &Fixed/[0,1]                  \\ 
		& log$g_{0}$                        & Cloud scattering asymmetry &
		log(0.8)                                & Fixed/[0,1]                 \\ 
		& log$f_{\rm cld}$                          & Cloud fraction                  & log(0.6)                             & {[}-2,0{]}       \\ \midrule                 
		\multirow{3}{*}{Surface albedo}                                                          & log$A_{b}$ (431–531 nm)                           & surface albedo of blue band     & log(0.09)                             & {[}-2,0{]}                       \\
		& log$A_{g}$ (569–693 nm)                         & surface albedo of green band    & log(0.09)                             & {[}-2,0{]}                       \\
		& log$A_{r}$ (770–894 nm)                           & surface albedo of red band      & log(0.18)                             & {[}-2,0{]}                       \\ \hline
		\label{tab1}
	\end{tabular}
\end{table*}

\subsubsection{MULTINEST nested sampling}

Ultimately, we are interested in extracting the underlying properties of an exoplanet atmosphere and surface from observed reflectance spectra. We then apply a Bayesian inference tool on the synthetic data set to sample the posterior probability distributions of the forward model input parameters. Table 1 lists the parameters and their corresponding prior ranges of our fiducial model. To perform Bayesian parameter estimation, we utilize the multimodal nested sampling algorithm MULTINEST \citep{feroz2008,feroz2013,feng2018} via the Python wrapper PYMULTINEST \citep{buchner2014x}. Briefly, the basic process includes 1) drawing a large number ($\sim$ 50$\times$ number of free parameters) of live points from the provided priors of the parameter space, and then 2) iteratively replacing the live point with the least likelihood with a new live point having a greater likelihood than the replaced point. In this manuscript, MULTINEST was run with 800 live points for each retrieval. The nested sampling should terminate when the expected evidence contribution from the current set of living points is less than a pre-defined tolerance (0.5). Usually, the final sample numbers are at the level of 1 million, which is determined by the sampling efficiency (0.3), parameter numbers, and the prior ranges.

\subsubsection{Fixed parameters}

Below, we list the parameters that are fixed in our retrieval study.

(1) stellar spectrum 

The source spectrum is assumed to be known. It is generated with the ATLAS9 stellar atmosphere models \citep{sbordone2004atlas} via python package of \textit{pysynphot} \citep{stsci2013pysynphot}.
The input spectrum used in the paper is  generated with metallicity [M/H] = 0, temperature $T_{eff}$ = 5300 K, and gravity $log(g)$ = 4.437. 

(2) The radius of the planet ($R_{p}$), the orbital separation (r), and the phase angle ($\alpha$).

As \citet{damiano2020}, we retrieve from the geometric albedo spectrum. The observed quantity for a directly imaged exoplanet in reflected light at a given phase angle, $\alpha$, is the wavelength-dependent planet-to-star flux ratio. 

\begin{equation}
\frac{F_{p}}{F_{s}}=A_{g} \Phi(\alpha)\left(\frac{R_{p}}{r}\right)^{2}
\end{equation}

where $F_{p}$ is the flux reflected by planet, $F_{s}$ is the incident flux from star, $A_{g}$ is the geometric albedo, $\Phi(\alpha)$ is the phase function, $R_{p}$ is the radius of the planet, and $r$ is the orbital separation. Thus, we assume that $R_{p}$, $r$, and $\alpha$ are known a priori to convert the flux ratio to geometric albedo. We have known that observer-planet-star phase angle impacts retrievals of atmospheric abundances, cloud properties, and there are degenerate relationships that exist between planet phase and planet radius \citep{nayak2017}. In this paper, we omitted this parameter degeneracy and focused on exploring the relationship between cloud properties and surface albedo.

(3) cloud compositions and their optical properties for the fiducial model and model I.

We used water clouds as on Earth. Earth's water cloud has a characteristic value of single scattering albedo $\omega$ = 1 and cloud scattering asymmetry $g_{0}\approx0.8$ across the visible range, which depends on the particle size. As \citet{feng2018}, we first fixed the cloud optical properties to evaluate the effectiveness of our surface albedo parameterization and facilitated the comparison with the gray albedo retrieval. After that, we retrieve with the full forward model, adding in the cloud properties $\omega$ and $g_{0}$ as fitting parameters. By doing so, we intend to explore the degeneracies between the cloud properties and surface albedo and figure out the information that needs to be constrained a priori in order to investigate the surface and cloud properties from reflected spectra of exoplanets.

\subsubsection{Experimental setup}

The key questions we ask in this work can be divided into:

(1) How do surface albedo parameterizations affect the inferred properties?

(2) How do the unknown cloud scattering properties affect what we can know about the atmosphere and the cloud?

(3) Could the multi-epoch observations help us eliminate the parameter degeneracies?

(4) Could the additional Rayleigh bands data help us eliminate the parameter degeneracies?

(5) Which parameters are critical to know a priori for a trustworthy parameter estimation?

Here we briefly describe the various surface, cloud parameterizations, and observation setups explored in this work.

\textbf{Surface parameterization}: 

\emph{Gray surface albedo} ($S_{\rm gray}$): The surface albedo is characterized with a wavelength-independent albedo of $A_s$.

\emph{The three photometric bands albedo} ($S_{\rm 3bands}$): The surface albedo is characterized with three photometric bands albedo of $A_{b}$, $A_{g}$, $A_{r}$.

\textbf{Cloud parameterization}: All cloud parameterizations are assumed to be a single deck water cloud with a homogeneous distribution. The variations are as follows:

\emph{Known cloud scattering properties case} ($C_{\rm \omega, g_{0}}$): We set the cloud single scattering albedo $\omega$ to 1, scattering asymmetry $g_{0}$ to 0.8. This is analogous to an Earth-like exoplanet with known cloud scattering properties in other retrieval studies (e.g., \citealt{lupu2016}).

\emph{A generalized cloud case} ($C$):  We fit simultaneously for the cloud opacity and scattering properties ($\omega$, $g_{0}$), but take both to be constant with wavelength.

\emph{Having estimates of cloud fraction} ($C_{\rm f_{cld}}$): We fit for the cloud opacity and scattering property ($\omega$, $g_{0}$), but we assumed partial constraints on cloud fractions instead. This is based on the expectation that cloud fraction may be independently estimated from the direct imaging observation. Future direct-imaging observations may allow us to obtain exoplanet cloud maps from time variations that constrain the exoplanet cloud and atmospheric circulation models \citep{cowan2008,fujii2012,cowan2013,fan2019,gu2021}. If these techniques could constrain cloud cover to a satisfactory level, this additional cloud cover information may help us tighten the retrieval constraint.
	
\begin{table*}[]
	\centering
	\caption {Summary of the Observational and Model Scenarios Explored }
	\begin{tabular}{@{}lllllll@{}}
		\toprule
		\toprule
		Experimental model & Model name & Npara\tablenotemark{a} & \begin{tabular}[c]{@{}l@{}}Surface \\ parameterizations\end{tabular} & \begin{tabular}[c]{@{}l@{}}Cloud \\ parameterizations\end{tabular} & Bandpass &  S/N \\ \midrule
		Fiducial model & $C_{\rm \omega, g_{0}}S_{\rm 3bands}$ & 12 & \begin{tabular}[c]{@{}l@{}}\multirow{1}{*}{$A_{b}$, $A_{g}$, $A_{r}$} \end{tabular} & \begin{tabular}[c]{@{}l@{}}\multirow{1}{*}{$p_{b}$, $d_{p}$, $\tau$, $f_{\rm cld}$} \end{tabular} & \multirow{1}{*}{0.5-0.9 $\mu$m} & 20 \\ \midrule
		I  & $C_{\rm \omega, g_{0}}S_{\rm gray}$ & 10 & \begin{tabular}[c]{@{}l@{}}\multirow{1}{*}{$A_{s}$} \end{tabular} & \begin{tabular}[c]{@{}l@{}}\multirow{1}{*}{$p_{b}$, $d_{p}$, $\tau$, $f_{\rm cld}$} \end{tabular} & \multirow{1}{*}{0.5-0.9 $\mu$m} & 20 \\
		II & $CS_{\rm 3bands}$ & 14 & \begin{tabular}[c]{@{}l@{}}\multirow{1}{*}{$A_{b}$, $A_{g}$, $A_{r}$} \end{tabular} & \begin{tabular}[c]{@{}l@{}}\multirow{1}{*}{$\omega$, $g_{0}$, $p_{b}$, $d_{p}$, $\tau$, $f_{\rm cld}$} \end{tabular} & \multirow{1}{*}{0.5-0.9 $\mu$m} & 20 \\
		III & $CS_{\rm 3bands}\_2phases$\tablenotemark{b} & 15 & \begin{tabular}[c]{@{}l@{}}\multirow{1}{*}{$A_{b}$, $A_{g}$, $A_{r}$} \end{tabular} & \begin{tabular}[c]{@{}l@{}}\multirow{1}{*}{$\omega$, $g_{0}$, $p_{b}$, $d_{p}$, $\tau$, $f_{\rm cld}$\tablenotemark{c}} \end{tabular} & \multirow{1}{*}{0.5-0.9 $\mu$m} & 20 \\
		IV & $CS_{\rm 3bands}\_sw$\tablenotemark{d} & 14 & \begin{tabular}[c]{@{}l@{}}\multirow{1}{*}{$A_{b}$, $A_{g}$, $A_{r}$} \end{tabular} & \begin{tabular}[c]{@{}l@{}}\multirow{1}{*}{$\omega$, $g_{0}$, $p_{b}$, $d_{p}$, $\tau$, $f_{\rm cld}$} \end{tabular} & \multirow{1}{*}{0.3-0.9 $\mu$m} & 20 \\
		V & $C_{\rm f_{cld}}S_{\rm 3bands}$ & 14 & \begin{tabular}[c]{@{}l@{}}\multirow{1}{*}{$A_{b}$, $A_{g}$, $A_{r}$} \end{tabular} & \begin{tabular}[c]{@{}l@{}}\multirow{1}{*}{$\omega$, $g_{0}$, $p_{b}$, $d_{p}$, $\tau$, $f_{\rm cld}$\tablenotemark{e}} \end{tabular} & \multirow{1}{*}{0.5-0.9 $\mu$m} & 20 \\
		\bottomrule 
	\end{tabular}
	\leftline{\textbf{Notes.}} 
	\leftline{\footnotesize{$^{\rm a}$ variable parameters + 5 constant parameters across all models (3 for atmospheric abundances and 2 for bulk properties).}}
	\leftline{\footnotesize{$^{\rm b}$ $2phases$ mean that the two spectra generated at different phase angles are used together to make a joint-retrieval.}}
	\leftline{\footnotesize{$^{\rm c}$ Here, the parameter $f_{\rm cld}$ represents the $f_{{\rm cld},1}$ for phase 1 and $f_{{\rm cld},2}$ for phase 2, while other parameters are all the same.}}
	\leftline{\footnotesize{$^{\rm d}$ $sw$  represents that the model uses additional UV band low-resolution spectra data (0.3-0.5 $\mu m$, R = 6) for retrieval.}}
	\leftline{\footnotesize{$^{\rm e}$ with partial constraints on $f_{\rm cld}$.}}
\end{table*}

\textbf{Bandpass}: We considered the following two bandpasses motivated by the proposed specification of Extreme Coronagraph for Living Planetary System (ECLIPS) for LUVOIR  \citep{roberge2019,roberge2021}:

\emph{Visible (0.5-0.9 $\mu m$)}: visible band spectra data with R = 140, which is the default observation bandpass.

\emph{Visible $+$ UV (0.3-0.9 $\mu m$, denoted as $sw$)}: We add the UV band for the consideration of Rayleigh scattering in this band. The spectra resolution of ECLIPS is utilized for UV: R $\sim$ 6, capable of characterizing cloud/haze scattering slopes.

\textbf{Observational phase}: The observational phase of $\alpha=90^{\circ}$ is utilized to generate the model spectra for all single-phase retrieval models. An additional phase angle of $\alpha=45^{\circ}$ is used together with $\alpha=90^{\circ}$ to make multi-epoch joint-retrievals.

The model in our experiment consists of different combinations of surface, cloud parameterization, and observational bandpass. We designed five models of retrieval experiments to explore the above questions accordingly. Firstly, we set up the fiducial model with the three photometric bands albedo $+$ known cloud scattering properties case $+$ visible bandpass ($C_{\rm \omega, g_{0}}S_{\rm 3bands}$). The retrieved parameters and prior ranges for the $C_{\rm \omega, g_{0}}S_{\rm 3bands}$  model are listed in Table 1.

In Experiment model I, we explore the role of surface albedo parameterizations on the performance of retrieval. The $C_{\rm \omega, g_{0}}S_{gray}$ model (gray surface albedo $+$ known cloud scattering properties $+$ visible bandpass) is compared with the fiducial mode of $C_{\rm \omega, g_{0}}S_{\rm 3bands}$.

In Experiment model II, we intend to examine a generalized retrieval scheme where all the cloud scattering properties ($\omega$, $g_{0}$), as well as surface spectral albedo, are treated as parameters to estimate. The model of  $CS_{\rm 3bands}$ with 3-bands surface albedo and a generalized cloud parameterization represents this scenario.

In Experiment model III, to overcome the difficulty with unknown cloud scattering properties, we use multi-epoch observations to make a joint-retrieval \citep{damiano2020multi,carrion2021}. This is because viewing the same atmosphere at different phase angles would produce a wavelength-dependent difference in the spectra as the illumination and emerging angles have chromatic effects in atmospheric absorption and cloud scattering \citep{cahoy2010,  madhusudhan2012, nayak2017}. \citet{carrion2021} concluded that physical reasons for the improvement in multi-phase retrievals are associated with the shape of the scattering phase function of the cloud aerosols.

Under the setup of $CS_{\rm 3bands}$ model, we calculate the spectrum for the first phase angle ($\alpha=45^{\circ}$) and compute the likelihood $\mathcal{L}_{1}$. With the same set of parameters except for a different cloud fraction, we calculate the spectrum at the second phase angle ($\alpha=90^{\circ}$) and compute the likelihood $\mathcal{L}_{2}$. The product, $\mathcal{L}_{1} \times \mathcal{L}_{2}$ will give the total likelihood of the fitting parameters set, as
\begin{equation}
	\log (\mathcal{L})=\log \left(\mathcal{L}_{1} \times \mathcal{L}_{2}\right)=\log \left(\mathcal{L}_{1}\right)+\log \left(\mathcal{L}_{2}\right)
\end{equation}
The $CS_{\rm 3bands}\_2phases$ model represents this experiment.

\begin{figure*}[!htb]
	\centering
	\includegraphics[height=16cm,width=18cm]{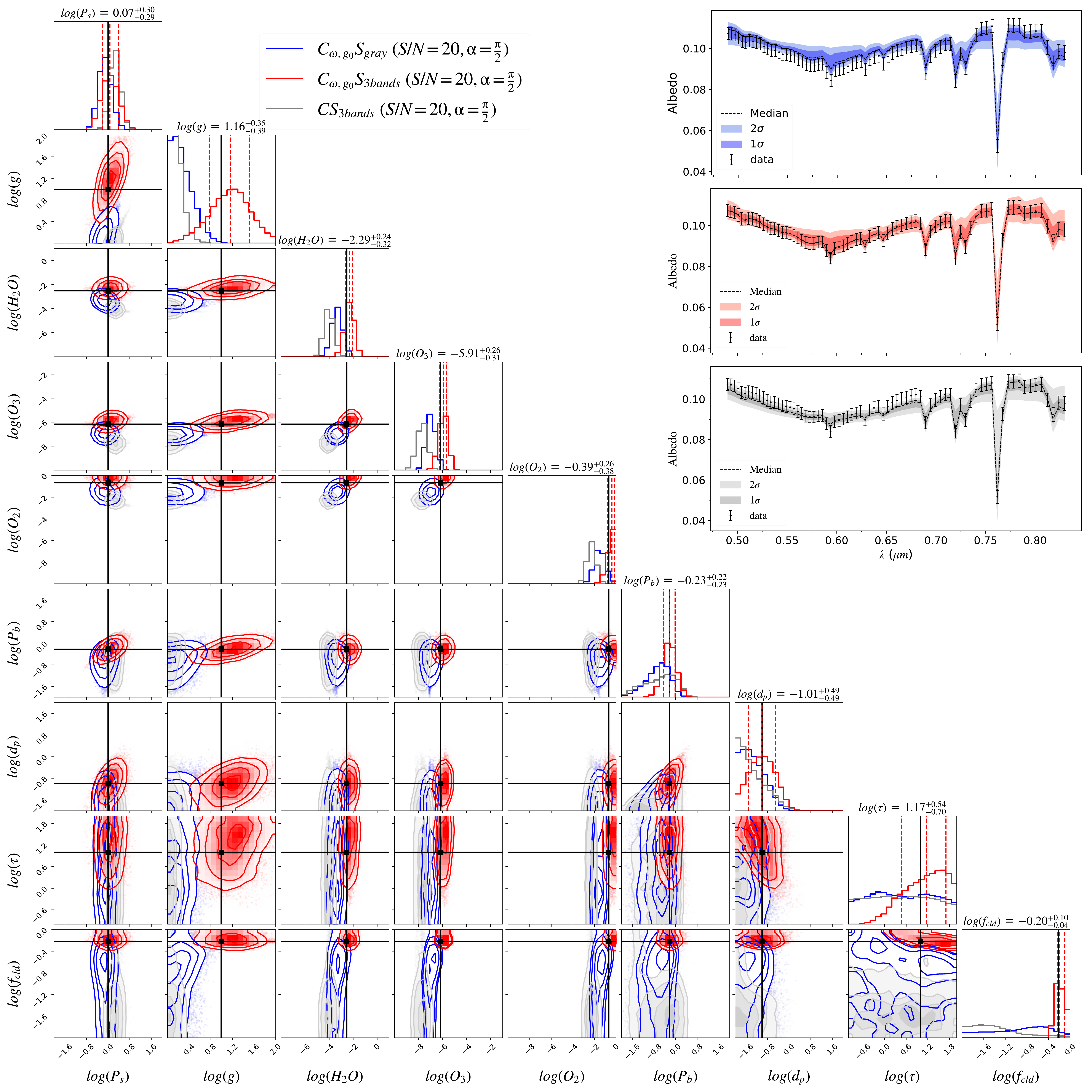}
	\caption{Posterior distributions of the common parameters from three retrieval schemes ($C_{\omega, g_{0}}S_{\rm gray}$, $C_{\omega, g_{0}}S_{\rm 3bands}$, $CS_{\rm 3bands}$; S/N = 20). Overplotted in black are the fiducial parameter values. Above the panel of 1D marginalized posterior distribution for each parameter, we indicate the median retrieved value with uncertainties of $C_{\omega, g_{0}}S_{\rm 3bands}$ model. Dashed lines (left to right) mark the 16\%, 50\%, and 84\% quantiles. Upper right panel: Spectra generated with 2000 randomly drawn parameters sampled from the retrievals. The color of the model spectrum is consistent with the color of contour shown on the left. Lighter contours represent 2$\sigma$ fits, while darker contours represent 1$\sigma$ fits. The solid line represents the median fit.}
	\label{img3}
\end{figure*}

In Experiment model IV, we start with the same surface and cloud parameterization of $CS_{\rm 3bands}$, but with a different bandpass.  The UV band of data that encoded the modulation of cloud are utilized together with the visible band data to see if the addition of the shortwave band can mitigate the difficulty with unknown cloud scattering properties ($CS_{\rm 3bands}\_{sw}$ model).

In Experiment model V, we examine the effect of possible additional prior information of cloud fraction on the improvement of parameter estimations. We set prior bounds on the parameters of $f_{\rm cld} \pm \delta f_{\rm cld}$ with varying relative uncertainty to discuss their effect on the elimination of parameter degeneracy. The $C_{\rm f_{cld}}S_{\rm 3bands}$ model adopted 5 different uncertainty levels of 30\%, 20\%, 10\%, 5\%, and 0\% (fixed $f_{\rm cld}$).

Table 2 summarizes the setup for the retrieval experiments described above.

\begin{figure}[!htb]
	\centering
	\includegraphics[height=4.8cm,width=8cm]{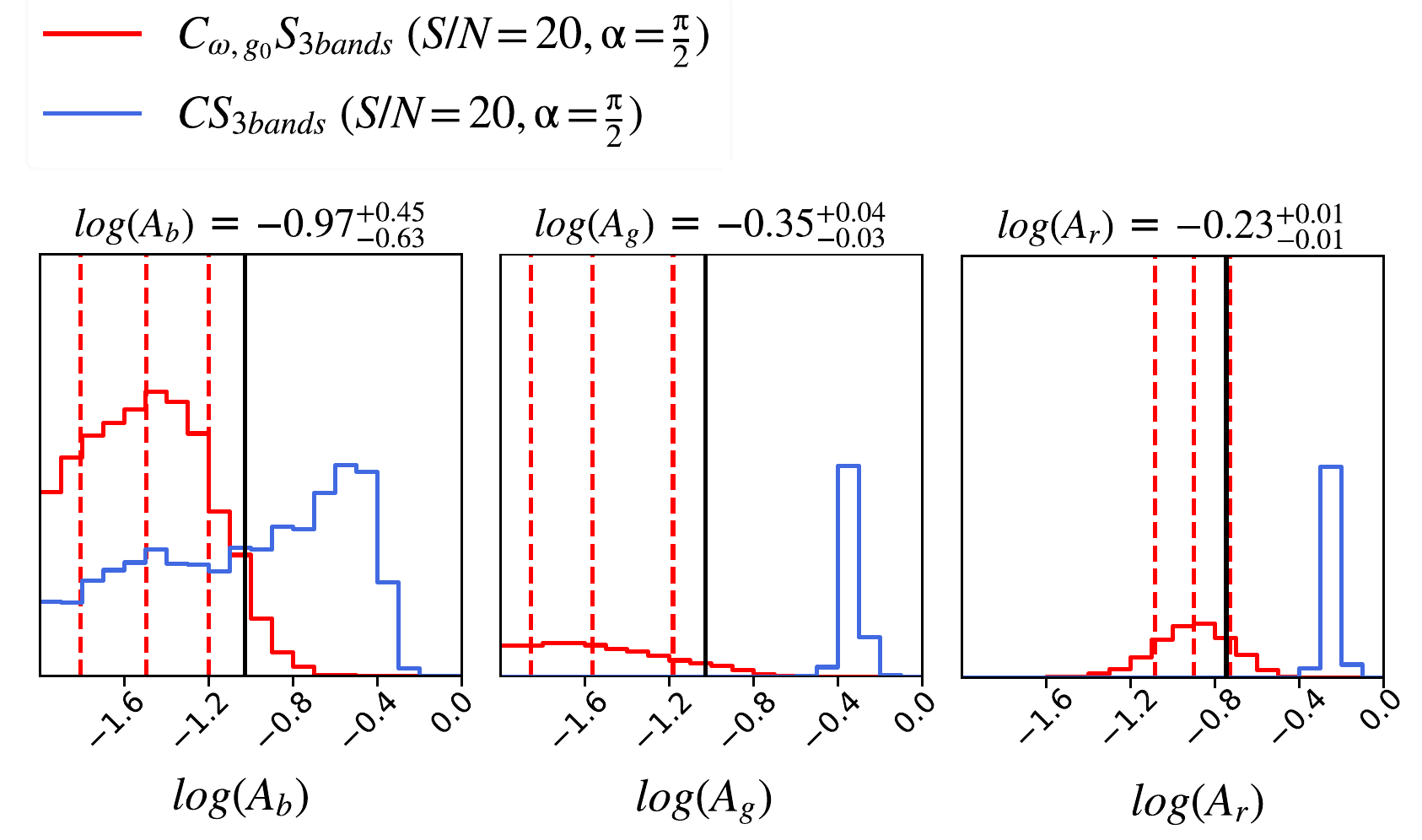}
	\caption{1D posterior probability distributions of surface albedo properties for model of $C_{\omega, g_{0}}S_{\rm 3bands}$ and $CS_{3bands}$. Overplotted in black are the fiducial parameter values. Above the panel of 1D marginalized posterior distribution for each parameter, we indicate the median retrieved value with uncertainties of $CS_{3bands}$ model with narrower prior range of surface gravity. The red dashed lines mark the 16\%, 50\%, and 84\% quantiles.}
	\label{img4}
\end{figure}

\section{Retrieval Results} 
Figures 3–6 illustrate the marginal posterior distributions of the key model parameters from each of the above experiments. Note that the parameters shown in Figure 3 are common among the $C_{\omega, g_{0}}S_{\rm gray}$, $C_{\omega, g_{0}}S_{\rm 3bands}$, $CS_{\rm 3bands}$ models. A higher probability corresponds to a darker shade in the two-dimensional correlation histogram. For each experiment, we randomly selected 2000 sets of parameter values from the remained set of live points (have reached to stop criterion) to calculate their corresponding spectra. The median spectrum and the 1$\sigma$ and 2$\sigma$ ranges are shown in the inserted panel of Figure 3.

\subsection{The effect of the wavelength dependence of surface albedos (fiducial vs. experimental model I)}

The red and blue contours of Figure 3 represent the results of the fiducial model ($C_{\rm \omega, g_{0}}S_{\rm 3bands}$ ) and experimental model I ($C_{\rm \omega, g_{0}}S_{\rm gray}$), respectively. The only difference between these two models is the albedo parameterization. As shown by the red colors in Figure 3, the $C_{\rm \omega, g_{0}}S_{\rm 3bands}$ model could constrain the key atmosphere parameters (H$_{2}$O, O$_{2}$, O$_{3}$) and cloud parameters ($P_{b}$, $d_{p}$, $\tau$, $f_{\rm cld}$) at the S/N of 20 in the sense that the true input values are within 1 $\sigma$ contours (i.e., no bias) and the values are constrained within less than one order of magnitude. For the surface albedo, as shown by the red histograms of Figure 4, the fiducial model could reasonably infer the $A_{r}$. At the same time, an upper limit on the $A_{b}$ and the $A_{g}$ can be obtained. The Rayleigh scattering worsens the constraints on albedo at shorter wavelengths because its efficient scattering at shorter wavelengths masks the surface. Besides, the albedos of the blue and green bands are close to zero; the retrieval can only give an upper limit. The red curves in the upper right of Figure 3 show the resulting median spectrum fit to the true spectrum.

\begin{figure}[!htb]
	\centering
	\includegraphics[height=8cm,width=8cm]{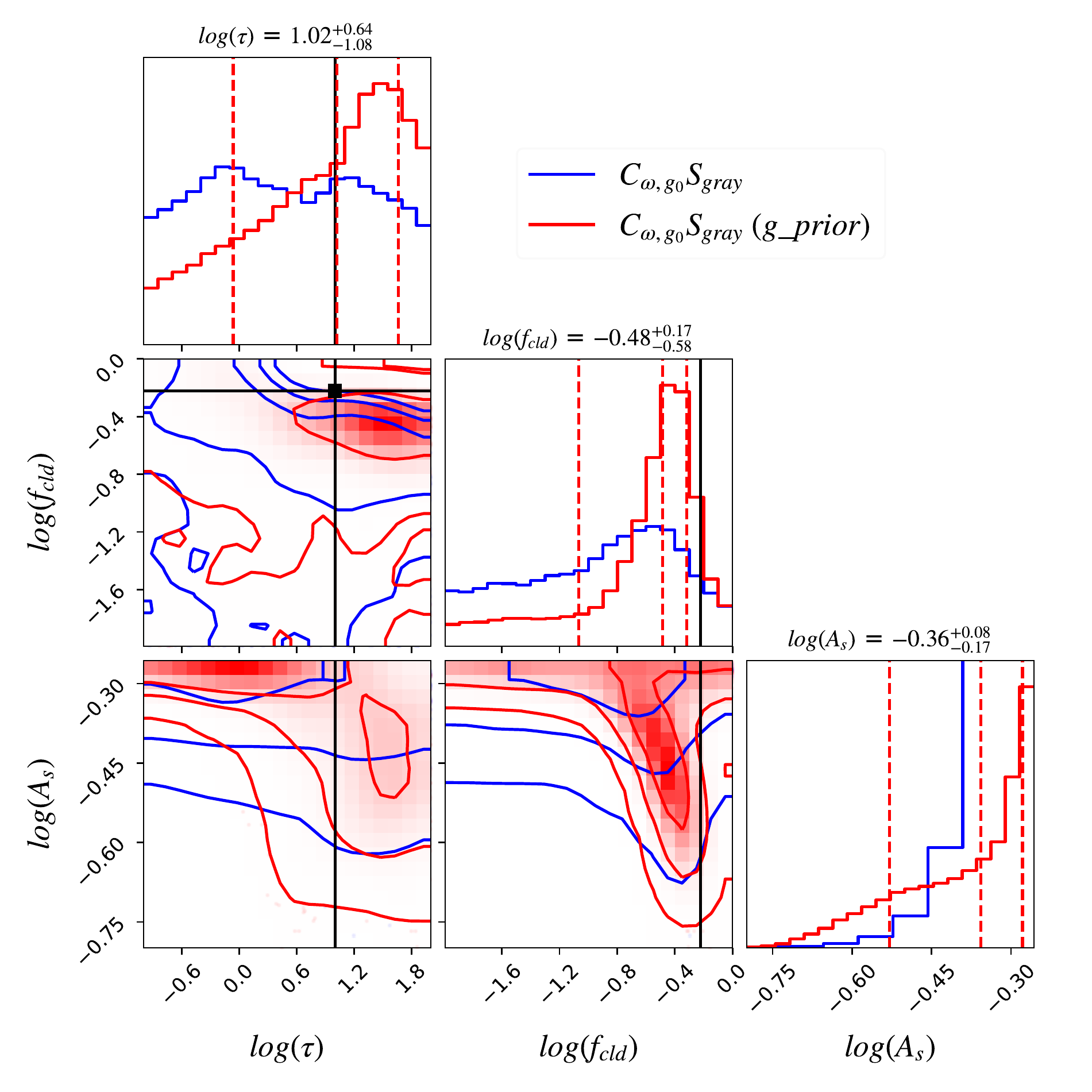}
	\caption{2D marginal posterior probability distributions of cloud optical properties ($f_{\rm cld}$, $\tau$) and surface reflectivity ($A_{s}$) for model of $C_{\omega, g_{0}}S_{\rm gray}$. The red contours represents the $C_{\omega, g_{0}}S_{\rm gray}$ model with narrower prior range of surface gravity (log$g \in$ [0.5, 1.5]). This retrieval experiment is given to show the degeneracy between surface albedo and cloud properties more clearly. Overplotted in black are the fiducial parameter values. Above the panel of 1D marginalized posterior distribution for each parameter, we indicate the median retrieved value with uncertainties of $C_{\omega, g_{0}}S_{\rm gray}$ model with narrower prior range of surface gravity. The red dashed lines mark the 16\%, 50\%, and 84\% quantiles.}
	\label{img5}
\end{figure}

The $C_{\rm \omega, g_{0}}S_{\rm gray}$ model (blue contours, Figure 3), where the retrieval spectrum is generated with gray surface albedo, results in biases. The key parameters of cloud properties ($\tau, f_{\rm cld}$) are unconstrained as the model presents a posterior distribution that does not contain information relative to the initially assumed prior. The estimation of surface gravity is $\sim$ 0.7 index less than the input value, and the bias of H$_{2}$O/ O$_{2}$/ O$_{3}$ mixing ratio is approximately an order of magnitude. The inferred gray surface albedo is $\sim$ 0.52, which is vastly larger than our input surface albedo spectrum. This bias can also be seen in the best fit spectrum in the upper right panel of Figure 3. The best fit spectrum (blue curves) deviates from the simulated data, which shows a relatively large albedo in the blue band.

These biases are simply due to our adoption of a gray surface albedo. In our ``ground truth" model, the albedo is generally more reflective in red than blue since land and plants' surface cover. In the case of gray surface albedo, to make the spectrum red and fit the high albedo value in the red band, O$_{3}$ absorption should be increased to reduce the reflectivity in the green band. Enlarging the O$_{3}$ absorption requires large column number density (i.e., smaller surface gravity) or large O$_{3}$ mixing ratios. The retrieval gives a degenerate solution biased to smaller surface gravity, which may be determined by minimizing the loglike function. The smaller surface gravity is probably the cause of the underestimated gas mixing ratios. Also, the lower value of $p_b$ and $d_{p}$ are made to compensate for the underestimated atmospheric mixing ratios.
	
We also show the parameter degeneracy between surface albedo and cloud properties in $C_{\rm \omega, g_{0}}S_{\rm gray}$ model in Figure 5. The blue and red contours represent results of $C_{\rm \omega, g_{0}}S_{\rm gray}$ model (log$g \in$ [0, 2]) and $C_{\rm \omega, g_{0}}S_{\rm gray}$ model with narrower prior range of surface gravity (log$g \in$ [0.5, 1.5]), respectively. The latter is given to show the degeneracy between surface albedo and cloud properties clearly. The parameters of $\tau$, $f_{\rm cld}$, and $A_{s}$ all control the overall spectral reflectance directly and thus making them degenerate with each other in Bayesian fitting processes. We emphasize that when the target planet's surface is wavelength-dependent, neglecting it in the retrieval process leads to a largely biased solution even for atmospheric parameters and makes the intrinsic degeneracy between parameters severer.

\subsection{The effect of unknown cloud scattering properties (fiducial v.s experiment model II)}

\begin{figure*}[!htb]
	\centering
	\includegraphics[height=16cm,width=18cm]{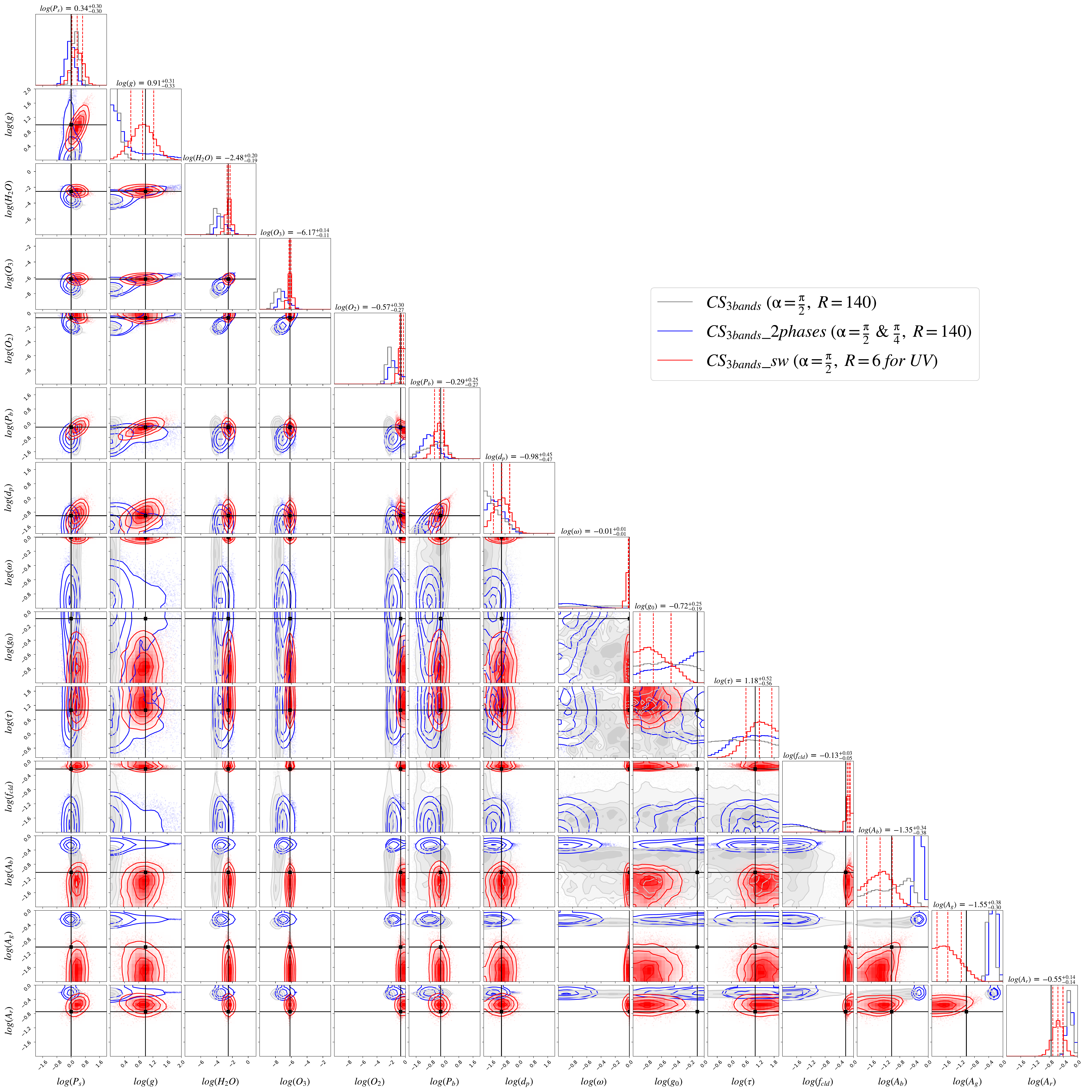}
	\caption{Posterior distributions of the model parameters from three retrieval schemes ($CS_{\rm 3bands}$, $CS_{\rm 3bands}\_2phases$, $CS_{\rm 3bands}\_sw$; S/N = 20). Overplotted in black are the fiducial parameter values. Above the panel of 1D marginalized posterior distribution for each parameter, we indicate the median retrieved value with uncertainties of $CS_{\rm 3bands}\_sw$ model. Dashed lines (left to right) mark the 16\%, 50\%, and 84\% quantiles. }
	\label{img6}
\end{figure*}

In this section, we retrieve with the full forward model, adding in the cloud scattering properties $\omega$ and $g_{0}$ as fitting parameters. This version of the model ($CS_{\rm 3bands}$ model) represents our most realistic scenario because we may not have a good guess about the cloud species and those optical properties in the visible band for a distant exoplanet. With this experiment, we try to examine a more generalized inversion framework to retrieve the cloud and the wavelength-dependent surface albedo properties simultaneously and intend to figure out what parameters may degenerate with each other in our model that finally influence the performance of retrieval.

The gray contours in Figure 3 and Figure 6 show the retrieval results of $CS_{\rm 3bands}$ model. We find that unknown cloud scattering properties can introduce significant biases. For cloud properties of $\omega$, $g_{0}$, and $\tau$, the $CS_{\rm 3bands}$ model presents flat posterior distributions that spread across the nearly entire prior range. This degenerate solution fits the simulated data spectrum relatively well, although there is an offset between the two spectra at shorter wavelengths, as shown in the upper panel of Figure 3 (gray curves).

Figure 4 also summarizes the surface albedo properties retrieved from the $CS_{\rm 3bands}$ model. The $CS_{\rm 3bands}$ model has almost no constraints on the surface albedo. This severe bias is intrinsic to the degeneracies between cloud properties of $\omega$, $\tau$, $f_{cld}$, $g_{0}$, and the surface spectral albedo. This is unsurprising given that both cloud properties and surface spectral albedo are correlated in the contribution of the total reflectance. If other atmospheric conditions remain unchanged, these two controlling reflectance factors can be compensated by each other. We will discuss this degeneracy in more detail in Section 4.1.

\subsection{Strategies to overcome the difficulty with unknown cloud scattering properties}

It comes as no surprise that the unknown cloud scattering properties result in difficulties in parameter estimation. Here, we intend to overcome this difficulty via three ways: 1) combining the multi-epoch observations to make a joint-retrieval; the $CS_{\rm 3bands}\_2phases$ model represents the joint-retrieval with two spectra simulated at the phase of $\frac{\pi}{4}$ and $\frac{\pi}{2}$. 2) adding another bandpass to bring more cloud information to the inverse retrieval process. We added a second wavelength bandpass: 0.3 – 0.5 $\mu m$ (UV band). The UV band is utilized together with the visible band for parameter estimations ($CS_{\rm 3bands}\_{sw}$ model). 3) assuming that the partial prior constraint on cloud fraction can be acquired ($C_{\rm f_{cld}}S_{\rm 3bands}$).

The reasons that we utilized the above three ways are as follows: For multi-epoch retrievals, the spectra observed at different phases reflect the shape of the scattering phase function of the cloud aerosols, which is determined by the illumination and emerging angles and cloud optical properties. For the additional UV band, from the slope of the Rayleigh scattering in this band, the information on the scale height of the atmosphere can be obtained, thus better constraining the surface gravity and future helping infer atmospheric mixing ratios. For the additional information on cloud cover fraction, cloud fraction itself is one of the main degeneracy sources; having its prior bounds may aid in excluding possible degenerate solutions.

Figure 6 allows quantitative comparisons of these different model scenarios. In addition, we clarify the simple criteria for preliminarily identifying whether a specific model retrieval is successful or not: its 1D marginalized posterior distributions of all model parameters should have peaked distributions. This means the correlations between all the model parameters are effectively explored in the retrieval process, even though the final result may bias from the true values.

\subsubsection{The effect of multi-epoch observations (experiment model III)}

\begin{figure}[!htb]
	\centering
	\includegraphics[height=6cm,width=8cm]{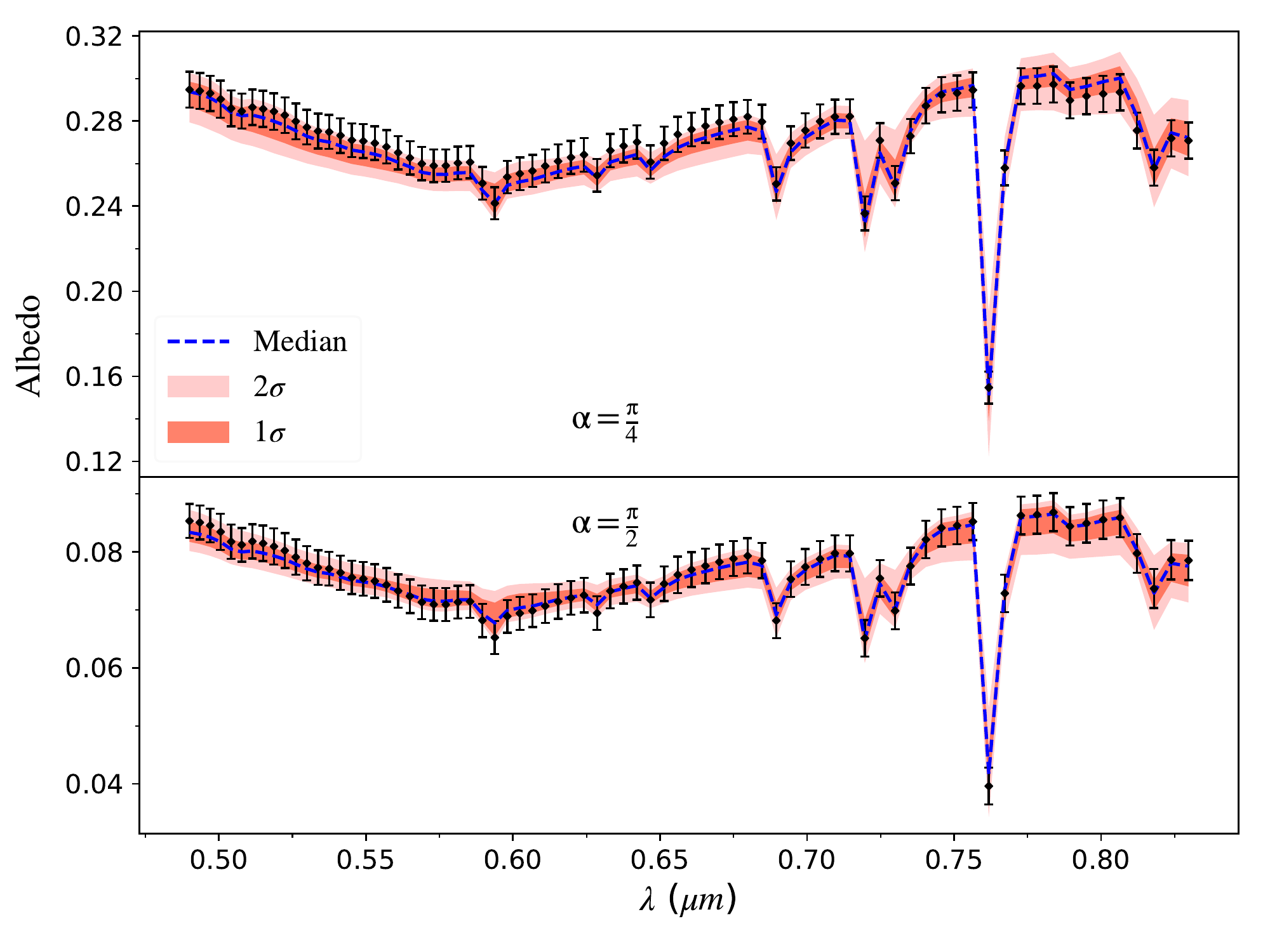}
	\caption{Spectra generated with 2000 randomly drawn sets of parameters sampled with the retrievals of $CS_{\rm 3bands}\_2phases$ model. Upper panel: retrieved spectra for $\alpha=\frac{\pi}{2}$. Lower panel: retrieved spectra for $\alpha=\frac{\pi}{4}$. The black dotted curves with error bars denote the reference data at S/N = 20; Lighter contours (tomato) represent 2$\sigma$ fits, while darker contours (red) represent 1$\sigma$ fits. The dashed blue curves represent the median fit.}
	\label{img7}%
\end{figure} 

As shown by \citet{carrion2021}, multi-epoch joint-retrieval could tighten the constraint on the atmospheric properties and the bulk properties in the case of known cloud scattering properties. A more thorough discussion of these retrieval improvements with the function of S/N can be found in Appendix A.

Here, we explored the effect of multi-epoch observations on parameter estimations under the case of unknown cloud scattering properties. Frustratingly, repeating the observation at a different orbital phase, the $CS_{\rm 3bands}\_2phases$ model seems to have little help on parameter estimation. The blue contours in Figure 6 show the retrieval results $CS_{\rm 3bands}\_2phases$ model. Compared with the unconstrained distributions of $CS_{\rm 3bands}$ model (gray contours), the parameters of $g_{0}$, $\tau$ now present peaked distributions, however with large uncertainties, and the medians are biased from the input values. The cloud single scattering albedo ($\omega$) is still unconstrained with its posterior distribution diffused all over its prior ranges. As shown in Figure 7, even though the retrieved spectra fit the data well at both two phases, we do not take the result of $CS_{\rm 3bands}\_2phases$ as a successful retrieval.

Adding another orbital phase observation is not informative enough to break all interactional degeneracy sources. The phase varied scattering phase function cannot pin down all cloud properties with wavelength-limited (0.5-0.9 $\mu$m) bandpass. The degeneracy between cloud properties and surface spectral albedo is still untangled, giving a local minimum solution.

\subsubsection{The effect of adding shorter wavelength (experiment model IV)}

Experiment model IV aims to explore whether the addition of the shortwave band can mitigate the difficulty with unknown cloud scattering properties.

As shown by the red contours of Figure 6, expanding the wavelength coverage to short wavelengths, the $CS_{\rm 3bands}\_sw$ model, when retrieved upon a spectrum with a bandpass of 0.3-0.9 $\mu m$, greatly improves the retrieval by determining the cloud fraction through Rayleigh scattering slope, although there remain some minor biases. Generally, the spectrum with a UV band (0.3-0.5 $\mu m$; R = 6) $+$ visible band (0.5-0.9 $\mu m$; R = 140) combination can reasonably constrain input values of atmospheric composition into 1 $\sigma$ contours under the case of unknown cloud scattering properties. The parameter of $g_{0}$ is still biased, thus leading to the offset of $f_{\rm cld}$ and surface spectral albedos, i.e., just outside the 1 $\sigma$ contours from this experiment. This biased solution also fits the data well, as shown by the red curve in Figure 8.

The adding feature -- the Rayleigh scattering slope in short bands, tightens surface gravity constraints. The relatively well-fitting values of $p_{s}$ and $g$ lead to a  relatively well-constrained total column density. The column densities of the radiatively active gases are also well-constrained from the depth of absorption bands. Besides, larger cloud cover reduces the average path length in the atmosphere, reducing the Rayleigh tail's relative intensity. This is probably the cause of the tighter constraint on cloud fraction and reduction of the retrieval degeneracy between cloud fraction and surface spectral albedo.

\begin{figure}[!htb]
	\centering
	\includegraphics[height=4cm,width=8cm]{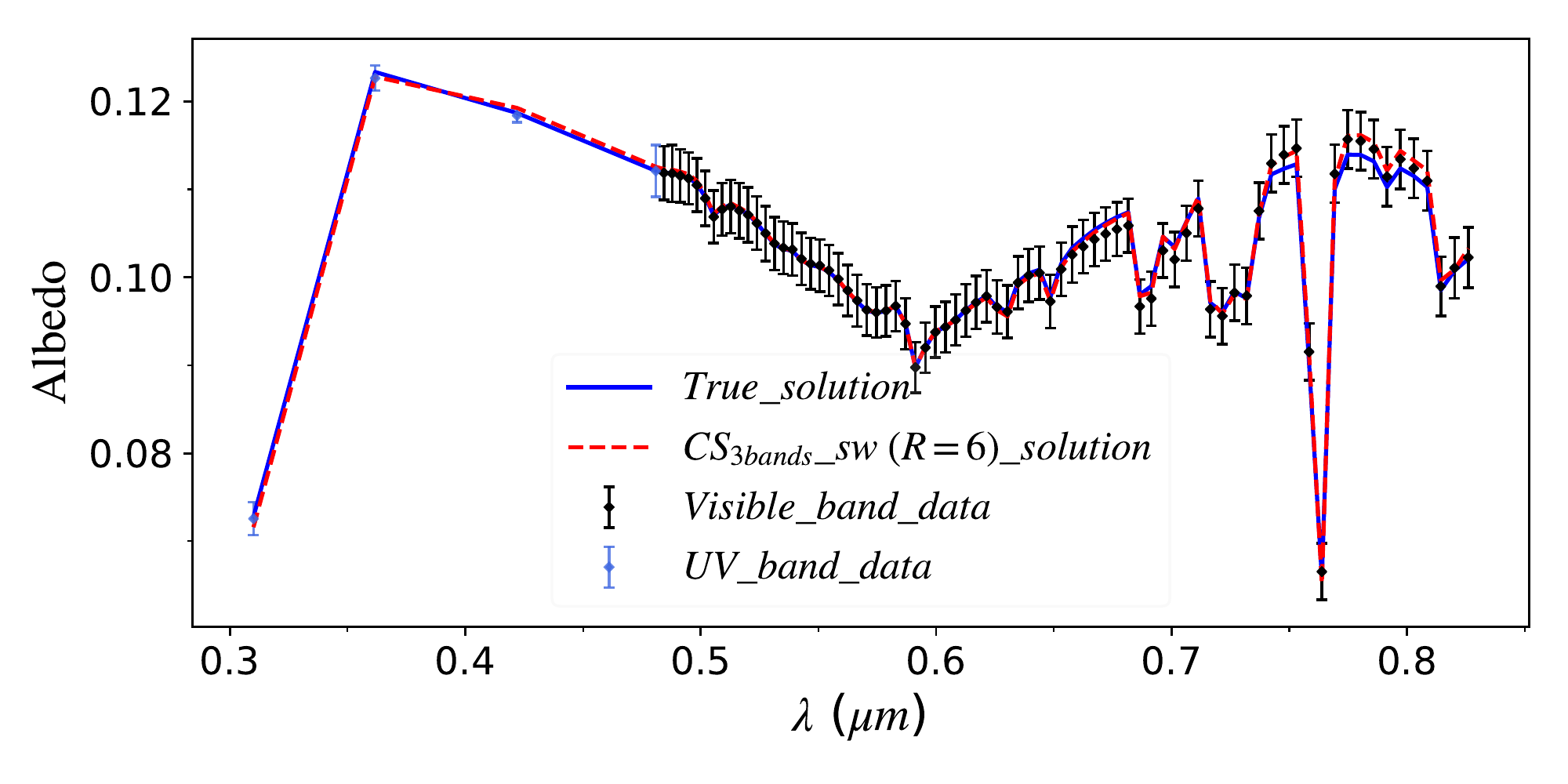}
	\caption{The spectrum of the degenerate solution of  $CS_{\rm 3bands}\_sw$ model is represented together with the simulated data. ``True$\_$solution" represents the spectrum generated with the input values. Dashed line for the degenerate solutions of $CS_{\rm 3bands}\_sw$ model and solid line for the true one. The spectral resolution of UV bands is set at 6.}
	\label{img8}
\end{figure}	
	
We note that the key improvement brought by the $CS_{\rm 3bands}\_sw$ model is related to the disentanglement between the surface albedo and cloud fraction. Thus, we may have to ask the following: if we cannot have short bandpass data, whether the additional information of cloud fraction can help to disentangle the degeneracy?

\subsubsection{The effect of having estimates of cloud fraction (experiment model V)}

\begin{figure*}[!htb]
	\centering
	\includegraphics[height=11cm,width=16cm]{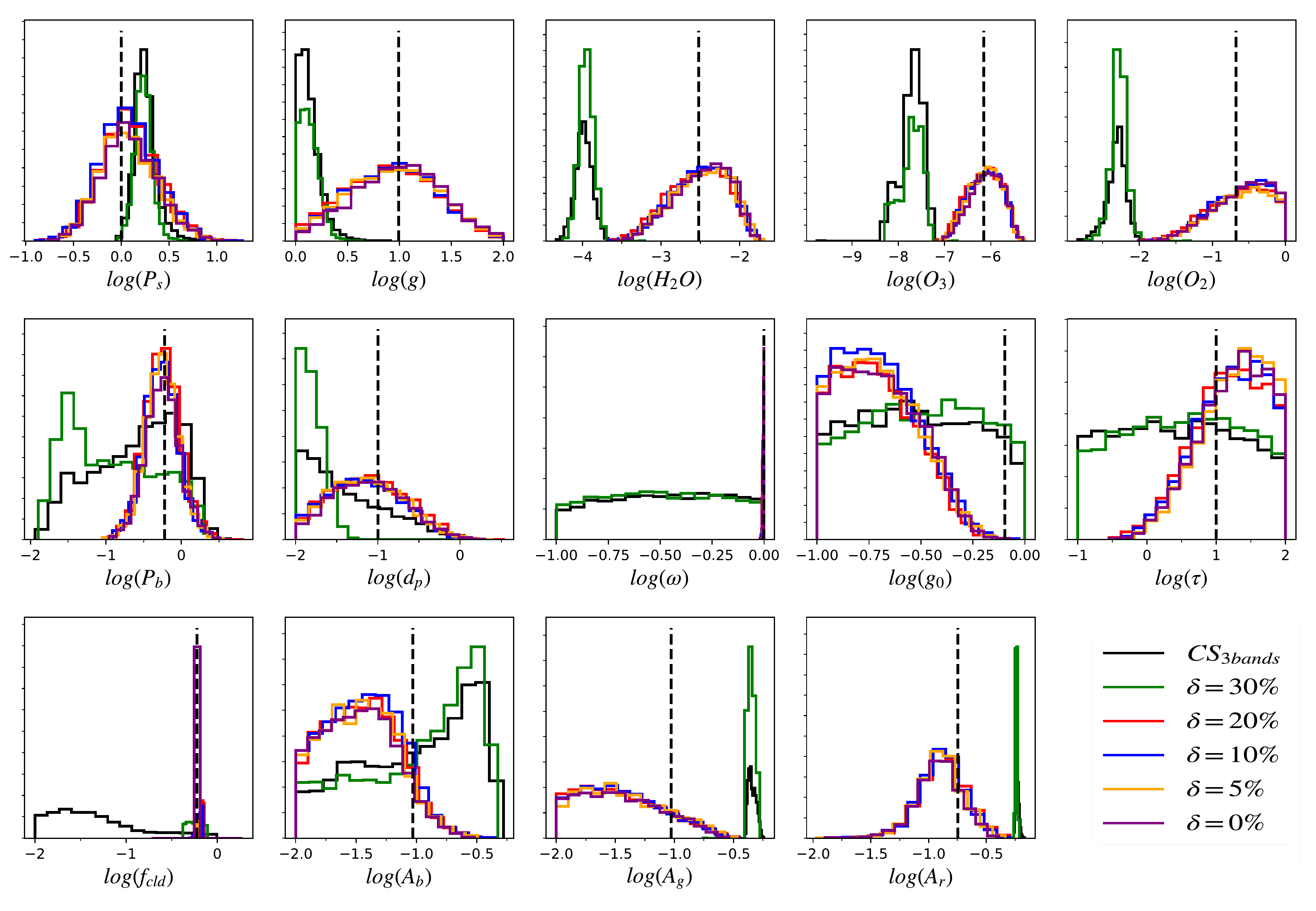}
	\caption{Summary of the 1D marginalized posterior probability distributions as a function of $f_{\rm cld}$ Relative Uncertainties ($\delta$). The dashed black lines represent the true values. Parameter constraints significantly improve as the uncertainty of the cloud fraction reduces from 30\% to 20\%. Further reducing the uncertainty does not change the 1 $\sigma$ confidence interval in a statistically significant manner.}
	\label{img9}
\end{figure*}

As yet another attempt to overcome the difficulty due to unknown cloud properties, in this section, we perform the retrieval with partially constrained cloud fraction  ($C_{\rm f_{cld}}S_{\rm 3bands}$).

Figure 9 shows the retrieval results of $C_{\rm f_{cld}}S_{\rm 3bands}$ models with different $f_{\rm cld}$ relative uncertainties ($\delta$). The $CS_{\rm 3bands}$ model is also given to represent the case of no information about $f_{\rm cld}$ (log$f_{\rm cld} \in$ [-2, 0]). For most key parameters, broader $f_{\rm cld}$  prior range increases the degenerancy complexity between cloud fraction and other parameter (e.g., $\omega$, $g_{0}$, and surface albedos). For example, a large uncertainty of $f_{\rm cld}$ ($\ge$ 30\%) leads to a lower retrieved atmospheric mixing ratios, lower $f_{\rm cld}$ but a more reflective surface. We can find that the parameter constraints significantly improve as the uncertainty of the cloud fraction reduces the from 30\% to 20\%. The $C_{\rm f_{cld}}S_{\rm 3bands}$ model with a relatively low $\delta$ ($\le$ 20\%) could include the input values of the bulk properties, atmospheric composition, and cloud properties to 1$\sigma$ contours. Moreover, the degenerate solutions that result in parameter biases (in experiments of $CS_{\rm 3bands}$ and the $C_{\rm f_{cld}}S_{\rm 3bands}$ with $\delta$ = 30\%) can be effectively eliminated. Tightening the $f_{\rm cld}$ prior range helps exclude the degenerate solutions with low $f_{\rm cld}$.

Another result is that further increasing the constraints on $f_{\rm cld}$ does not significantly improve the overall model parameter estimations. As shown in Figure 9, for most key parameters, the inferred values have reached true values at $\delta$ of 20\%. Further tightening the prior range of $f_{\rm cld}$ does not change the 1 $\sigma$ confidence interval in a statistically significant manner. Even with the strongest constraint on cloud fraction (fixed $f_{\rm cld}$), there are still biases on $g_{0}$, and the surface albedo of $A_{b}$, $A_{g}$. The true value lies outside the bounds of the  1$\sigma$ confidence interval. This may be due to the additional prior information about $f_{\rm cld}$ cannot fully break the degeneracy between surface albedo and cloud scattering properties. However, to place a preliminary constraint on cloud properties and surface spectral albedo, the partial constraint ($\delta$ $\le$ 20\%) on cloud fraction is critical to know or assume a priori.

\subsection{Placing constraint on surface composition under the case of known cloud scattering properties}

We have shown that the ``red" surface albedo can be constrained using the model of $C_{\rm \omega, g_{0}}S_{\rm 3bands}$, i.e., with the known cloud scattering properties. Here we examine whether other surface compositions that may be present on Earth-like planets can be constrained using the same framework. After all, constraining the surface composition is important in assessing the habitability of terrestrial exoplanets.

In particular, we intend to constrain the vegetated areas and ocean, two representative Earth surface covers. We note that the constraints on these two surface compositions are made by looking for the unique albedo feature that varied with wavelength (VRE) or phase (ocean glint) in our inferred three bands albedos. To verify the determined surface compositions, it is critical to cross-validate with other auxiliary observational data. This can be done by observing time variability photometry and spectra (e.g., \citealp{ford2001,wang2021}) to monitor the time-varied surface albedo modulated by the specific surface type.

\subsubsection{Vegetated areas}

\begin{figure}[!htb]
	\centering
	\includegraphics[height=8cm,width=8cm]{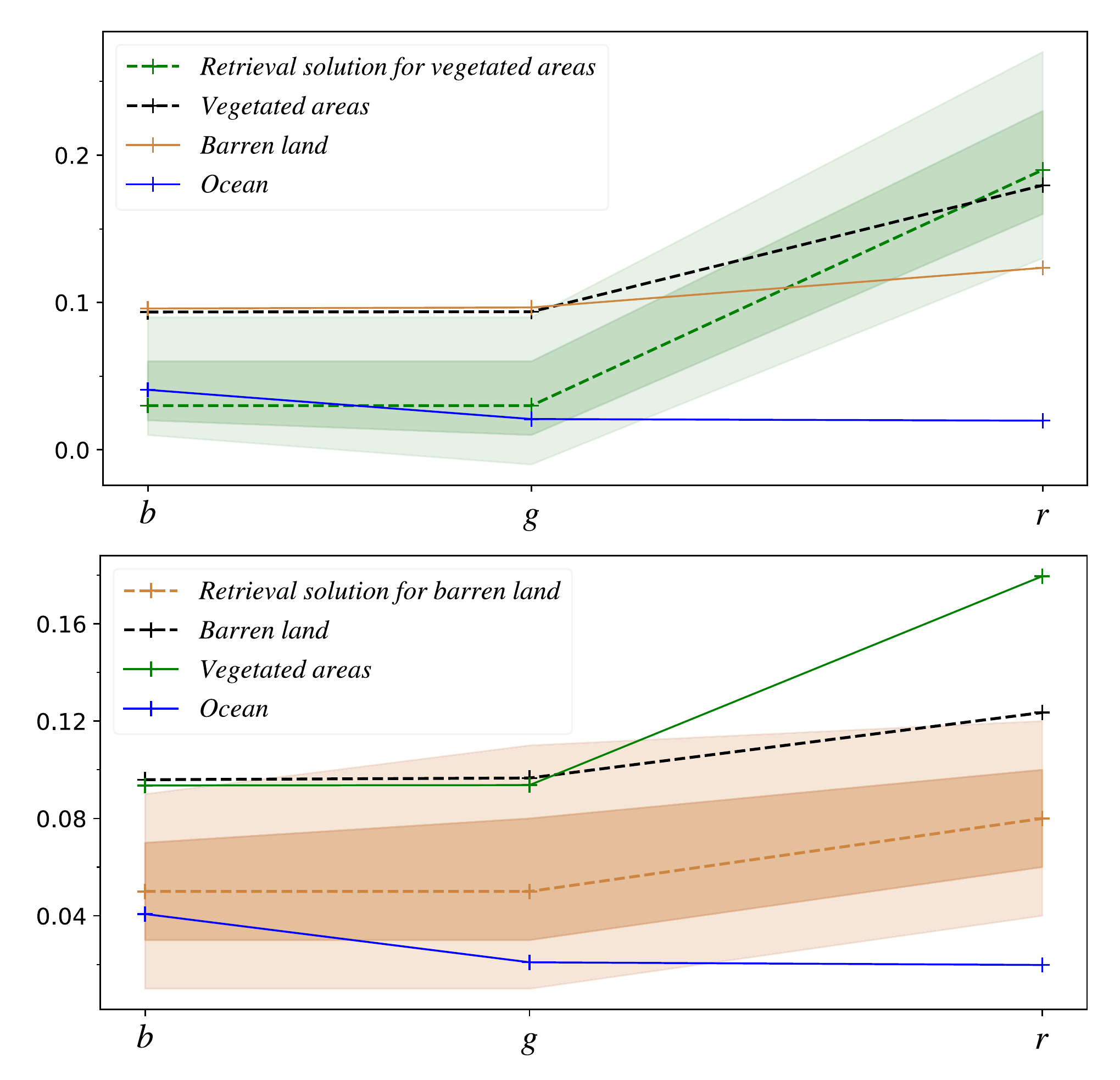}
	\caption{Comparison between retrieved albedos and the true values. The colored shaded areas indicate the 1$\sigma$ (darker contours) and the 2$\sigma$ (lighter contours). Upper panel: vegetated areas scheme. The black dashed curve stands for the truth. The inferred band albedos (green dashed curve) can discriminate the ocean (blue curves), but the barren land (brown curves) could also fit the retrieval. Lower panel: barren land scheme. The black dashed curve stands for the truth. The inferred band albedos (brown dashed curve) could eliminate the surface compositions of the ocean and the Earth-like vegetated areas.}
	\label{img10}
\end{figure}	

Earth-originated vegetated areas have their unique surface reflectance feature of vegetation red edge (VRE), which corresponds to a jump in reflectivity around 0.7 $\mu$m \citep{seager2005}. Previous works have answered the questions of how detectable the ``red edge” or other biological pigment-induced albedo features with the disk-integrated spectral model simulations \citep{hegde2015,walker2018,tinetti2006,montanes2006,o2018}. Here, we extend to determine whether future data can constrain vegetated areas through spectral retrievals given the observed data.

The upper panel of Figure 10 presents the median value of retrieved surface albedo of vegetated areas and corresponding 1$\sigma$, 2$\sigma$ fits. The surface albedo of the red band is well-constrained, while the upper limit can be obtained for albedo at shorter wavelengths ($A_{b}$ and $A_{g}$) where the absolute value of the input albedo is low, and Rayleigh scattering tends to mask the surface feature. A steep reflectance increase between the green and red bands can be clearly located (green dashed curve). This is encouraging; however, we cannot say that the retrievals are entirely diagnostic to vegetated areas given this level of fidelity of the retrieved band surface albedos and a wide range of different exoplanets surface covers \citep{hu2012}. 
\begin{figure}[!htb]
	\centering
	\includegraphics[height=4cm,width=8cm]{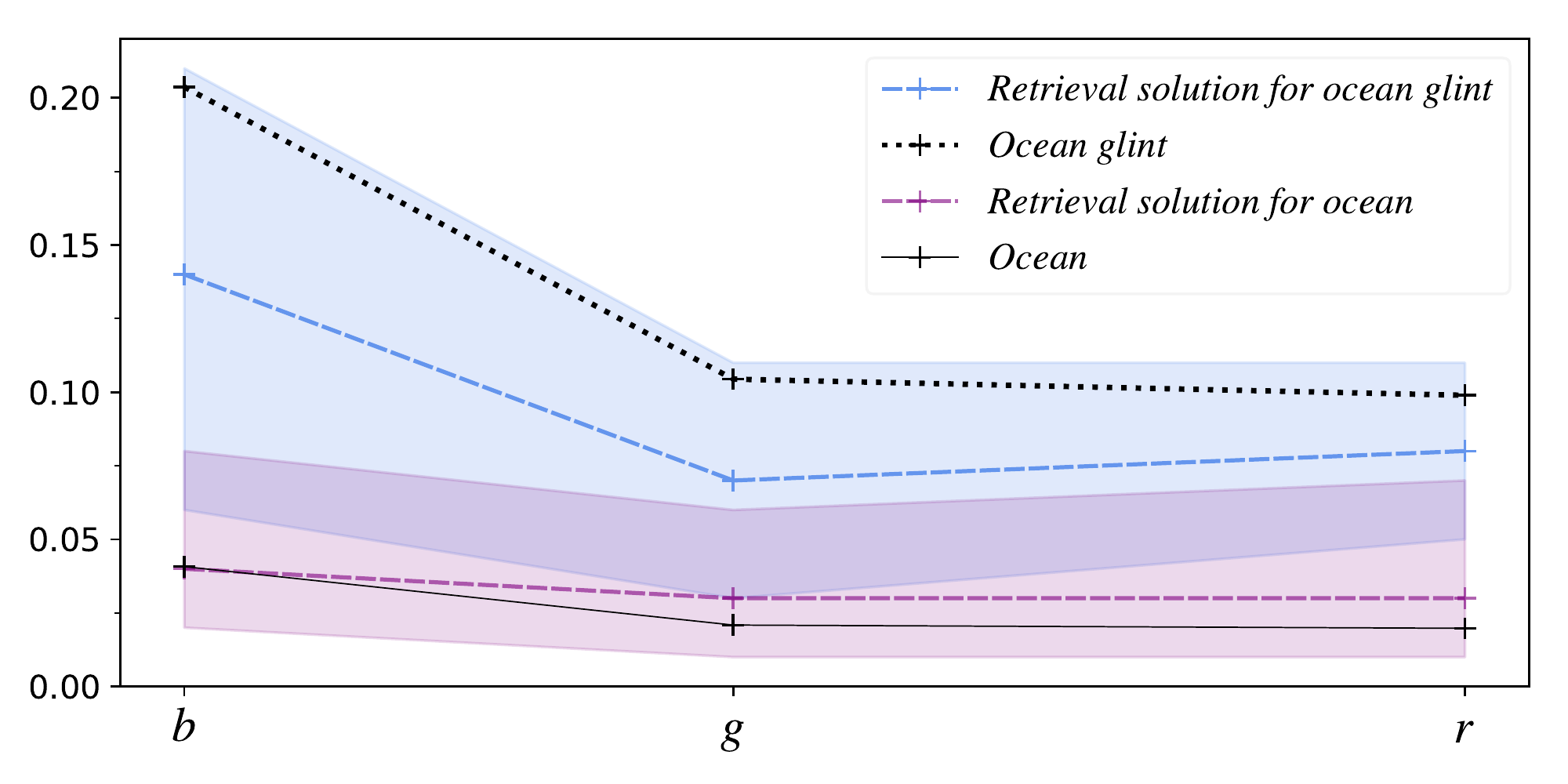}
	\caption{A comparison between retrieved ocean albedo to the truth at the phase of quadrature and crescent. The retrieved ocean glint albedo is shown with blue dashed curves compared to the true values (black dotted curves). Similarly, the inferred ocean albedo at the quadrature phase is given with purple dashed curves. The colored shaded regions indicate the 1$\sigma$ uncertainty envelopes for the retrieved albedo.}
	\label{img11}
\end{figure}
As shown in the upper panel of Figure 10, the water ocean can be effectively eliminated. However, the retrieval also includes the albedos of barren land in the 2$\sigma$ contours, although its slope (brown curve) between green and red is lower than the one brought by vegetation photosynthesis.
 
\subsubsection{Barren world}

For comparison, we computed retrieval with setup of $C_{\rm \omega, g_{0}}S_{\rm 3bands}$ model while with the surface cover of barren land. In this retrieval experiment, the vegetated areas of Earth are replaced with barren land while keeping other surface covers unchanged. As shown in the lower panel of Figure 10, the true values of barren land albedo can be restricted to 2$\sigma$ uncertainty envelopes.
Oceans (blue curve) and vegetated areas (green curve) similar to the Earth can be effectively excluded. Therefore, when the information given by retrieval is that the surface albedo has a considerable albedo uplift between the green and red bands, it is likely that there will be vegetated areas on the surface of the target exoplanet.

\subsubsection{Ocean} 
We also explore the possibility of constraining the ocean surface via retrieving the surface albedo difference between the phase of quadrature and crescent. A large ocean should appreciably have the typical specular reflection when viewed at an oblique angle relative to the illumination source. This glint effect produces a substantial relative increase in the brightness of Earth when viewed at crescent phases \citep{robinson2014}. Since the forward model used in this paper did not include the specular reflection of the ocean, we made a simplification to replace the specular reflection of the ocean with a relatively high surface albedo to mimic the glint effect at the phase of the crescent. The typical glint reflectance in the specular direction is on the order of 0.2, which is significantly smaller than, for instance, a typical cloud reflectance \citep{breon2006}. In \citet{lustig2018}, the retrieved apparent albedo of ocean-bearing longitudinal slices at the crescent phase increased by a factor of 5, compared to the albedo at the quadrature phase, due to the contribution from glint. Therefore, we simply multiply the ocean albedo with a factor of 5 within the model wavelength range.

Figure 11 shows the inferred value of ocean albedo from the model of $C_{\rm \omega, g_{0}}S_{\rm 3bands}$, which is consistent with the true value within the purple shadowed 1$\sigma$ areas. Similar results can be seen in the retrieved specular albedo schemes. The scale factor between the two phase angles of the blue band obtained by inversion is ${3.24}_{-1.98}^{+4.82}$, whose median value is smaller than the input value of 5. However, this unique phase-varied albedo is consistent with the effect of ocean glint, suggesting that the exoplanet may compose of large areas of the ocean-bearing surface.

\section{Discussion}
We investigated mitigated strategies in handling cases with completely unknown cloud scattering properties and surface spectral albedo. In the following subsections, we discuss the notable correlations/degeneracies within $CS_{\rm 3bands}$ model (Section 4.1), the requisite prior information to give a trustworthy retrieval for different parameters (Section 4.2), and the model limitations (Section 4.3).

\subsection{Notable parameter degeneracies} 

Previous studies have shown that there can be substantial degeneracies between the atmospheric composition, the cloud top pressure, surface gravity, and surface pressure (e.g., \citealp{lupu2016,nayak2017,hu2019,carrion2020}) when the wavelength coverage is limited (visible to near-infrared) and with moderate wavelength resolution and S/N.  Also, there is an expected correlation between $\tau$ and $f_{\rm cld}$ \citep{feng2018}; a higher cloudiness fraction can complement a less optically thick cloud and vice versa.

Here, we discuss the degeneracy between the cloud properties and the surface spectral albedo. This degeneracy will add to the uncertainties and have to be seen as a caveat when the cloud scattering properties cannot be estimated a priori. In $CS_{\rm 3bands}$ model, there are two main degeneracy sources: 1) degeneracy among cloud properties: $\omega$, $g_{0}$, and $f_{\rm cld}$. 2) degeneracy between cloud properties and surface albedo: $\omega$, $f_{\rm cld}$, and $A_{r}$. To see these degenerate sources more clearly, now we perform the retrieval, fixing other parameters at the fiducial values and adjusting the fitting parameters set of ($\omega$, $g_{0}$, $f_{\rm cld}$) or ($\omega$, $f_{\rm cld}$, $A_{r}$).
 
\begin{figure}[!htb]
	\centering
	\includegraphics[height=8cm,width=8cm]{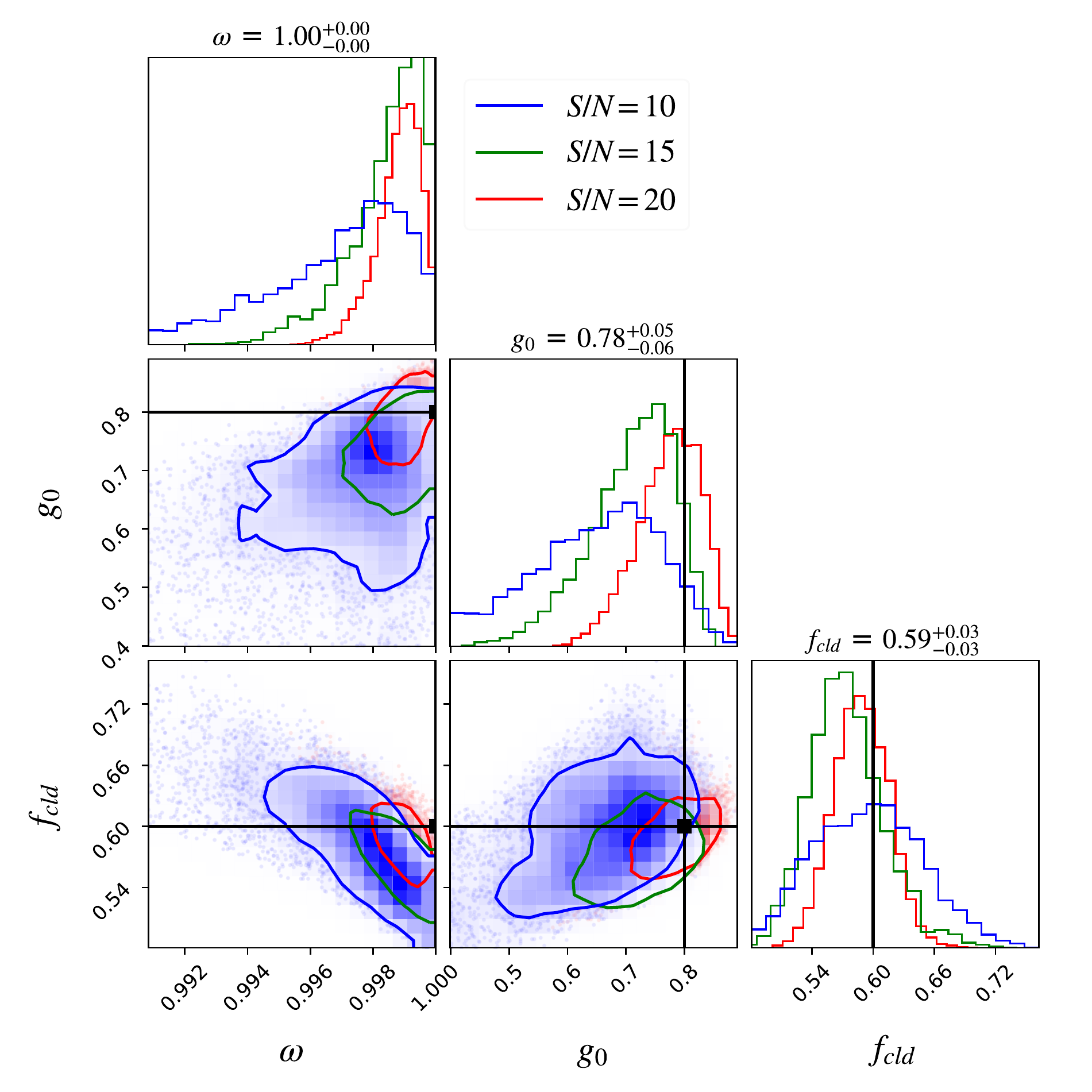}
	\caption{Response of the degeneracy among parameters of $\omega$, $g_{0}$, and $f_{\rm cld}$ to varying S/N. The colored curves represent the 68\% confidence intervals as a function of S/N, with blue representing the lower S/N data and red representing the higher S/N. The black crossed lines indicate the true value. Above the panel of 1D marginalized posterior distribution for each parameter, we indicate the median retrieved value in the case of S/N = 20.}
	\label{img12}
\end{figure}

\subsubsection{Degeneracy among cloud properties}

Figure 12 summarizes the source of the degeneracy among $\omega$, $g_{0}$, and $f_{\rm cld}$. We find that as $\omega$ increases (toward the right), $f_{\rm cld}$ must decrease. Moreover, we can compensate a larger $g_{0}$ with a larger $\omega$. This is unsurprising given that $\omega$, $f_{\rm cld}$, and $g_{0}$ are intrinsically correlated with the overall reflectance (Figure 2 panel a, c, and d). The former two parameters affect the reflectance of spectra positively, while $g_{0}$ negatively. Therefore, if other atmospheric conditions remain unchanged, to fit a spectrum, we can trade a larger $g_{0}$ with a larger $\omega$ or $f_{\rm cld}$ and maintain a similar overall albedo.

Figure 12 also explores how increased signal to noise influences this degeneracy. The case S/N of 5 is omitted because the posterior distribution disperses within the relatively large prior ranges, and the correlation between parameters is not distinct. The improved S/N leads to a posterior more concentrated around the true value (red curves). The degeneracy is largely “reduced” from the high-$f_{\rm cld}$, low-$\omega$, and low-$g_{0}$ end and reach to the fiducial value ($f_{\rm cld}$ = 0.6, $\omega$ = 1, $g_{0}$ = 0.8), which means that higher S/N could narrower the solution space. 

\begin{figure}[!htb]
	\centering
	\includegraphics[height=8cm,width=8cm]{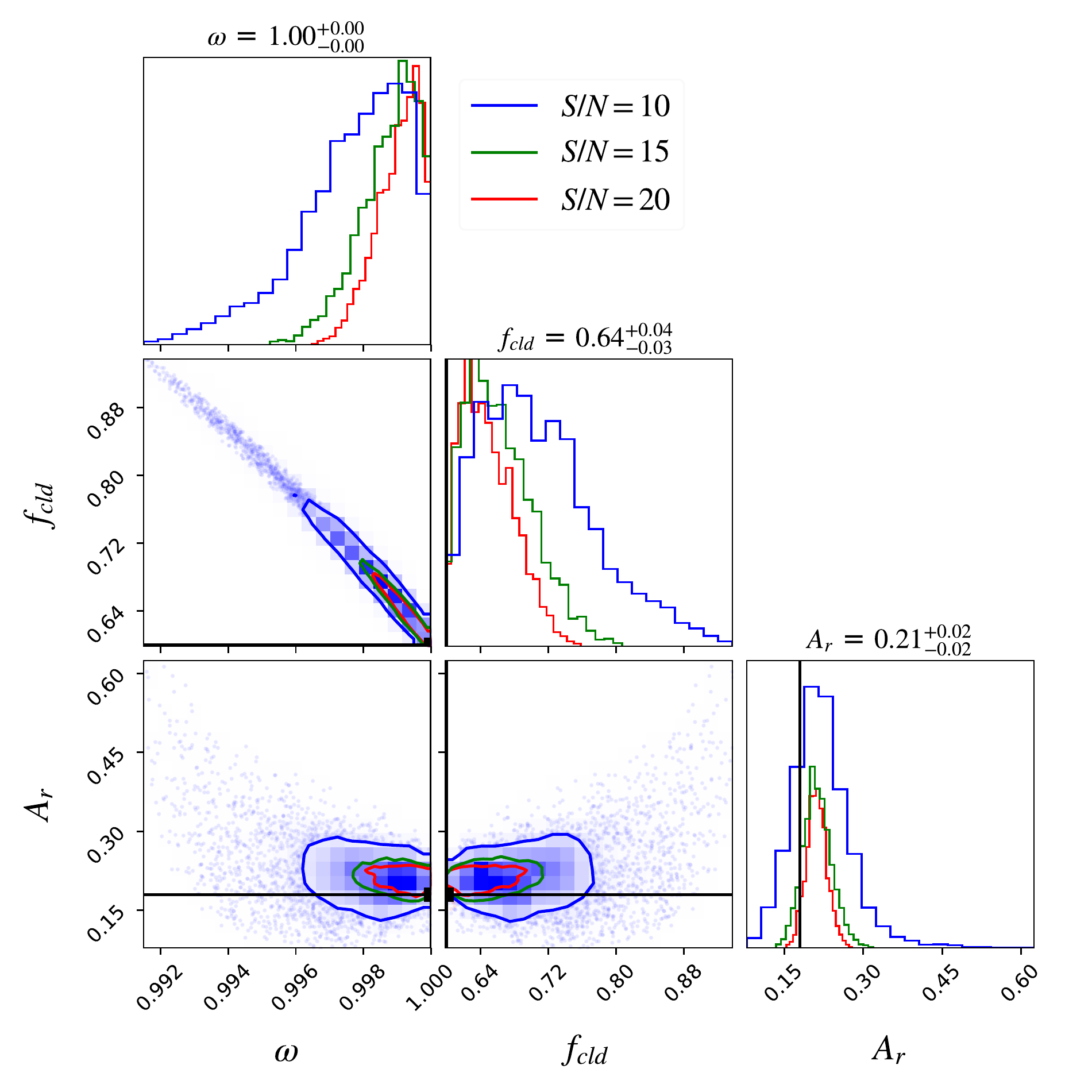}
	\caption{Response of the degeneracy among parameters of $\omega$, $f_{\rm cld}$, and $A_{r}$ with S/N. The colored curves represent the 68\% confidence intervals as a function of S/N, with blue representing the lower S/N data and red representing the higher S/N. The black crossed lines indicate the true value. Above the panel of 1D marginalized posterior distribution for each parameter, we indicate the median retrieved value in the case of S/N = 20.}
	\label{img13}
\end{figure}

\subsubsection{Degeneracy between cloud properties and surface albedo}

Degeneracy also exists between cloud properties and surface spectral albedo. We have shown the degeneracy between the gray surface albedo of $A_{s}$ and cloud properties of $f_{\rm cld}$, $\tau$ in Figure 5. All of these parameters directly control the overall reflectivity across the whole bandpass, naturally causing the degeneracy. Here we explore the degeneracy among $\omega$, $f_{\rm cld}$, and surface albedo of  red band ($A_{r}$). 

As we mentioned before, the larger the $\omega$ is, the lower $f_{\rm cld}$ is. Notably, this degeneracy between $\omega$ and $f_{\rm cld}$ is tighter in Figure 13 than in Figure 12. This is because of the relatively less prominent degeneracy between the $A_{r}$ and $f_{\rm cld}$. The model spectrum has a distinct wavelength dependence in surface albedo, which can only be represented by $A_{r}$, as $f_{\rm cld}$ increases the reflectivity at all visible wavelengths. This implies that the degeneracy between cloud properties and surface spectral albedo can be reduced via using 3-bands surface albedo parameterization if the surface albedo has a distinct spectral shape. On the other hand, when the albedo difference is not distinct, the cloud can easily compensate for the surface reflectance. The 3-bands surface albedo parameterization may degrade to gray surface parameterization, offering limited help in reducing this degeneracy. 

\subsection{Prerequisite information for a trustworthy parameter inference}

A reflected-light spectrum manifests itself with two main features: the shapes of molecular absorptions and its overall spectra reflectance. Correspondingly, the degeneracies encountered in the model of $CS_{\rm 3bands}$ are mainly composed of two types: 1) degeneracy that influences the shapes of molecular absorptions: degeneracy among bulk properties, atmospheric mixing ratios, cloud top pressure, cloud optical thickness, and cloud fraction. 2) degeneracy that influences the overall spectra reflectance: degeneracy between cloud scattering properties and surface albedo. These two degeneracies are entangled through the parameter of cloud optical thickness and cloud fraction, making the retrieval more complicated. Thus, it is important to clarify the degeneracy sources and provide additional information to obtain a trustworthy parameter estimation. 

\begin{table*}[!htb]
	\caption {Retrieved parameters for different prior information scenarios. The parameters that are usefully constrained are checked.}
	\begin{tabular}{@{}cccccc@{}}
		\toprule
		\toprule
		Fixed & Bulk properties & gas mixing ratio & Cloud vertical distribution  &  Cloud optical properties \& fraction & Surface spectral albedo \\ \midrule
		
		g & \XSolidBrush    & \XSolidBrush & \XSolidBrush  & \XSolidBrush & \XSolidBrush \\
		$g\ \&\ P_{s}$ & \textbf{--} & \CheckmarkBold & \CheckmarkBold & \XSolidBrush & \XSolidBrush \\ 
		
		$\omega\ \&\ g_{0}$ & \CheckmarkBold & \CheckmarkBold & \CheckmarkBold & \CheckmarkBold & \bm{$\circleddash$}\\
		
		$f_{\rm cld}$ &  \CheckmarkBold& \CheckmarkBold & \CheckmarkBold & \bm{$\circleddash$} & \bm{$\circleddash$}
		\\ \bottomrule
		\label{tab3} 
	\end{tabular}
	\leftline{\textbf{Notes.}}
	\leftline{cloud optical properties \& fraction: $\tau$, $\omega$, $g_{0}$, $f_{\rm cld}$; cloud vertical distribution: $P_{b}$, $d_{p}$.}
	\leftline{\footnotesize{\XSolidBrush: Parameter unconstrained -- the true value lies outside the bounds of the 1 $\sigma$  interval or the 1 $\sigma$ interval is close to matching the input prior.}} 
	\leftline{\footnotesize{\CheckmarkBold: Parameter constrained -- the true value falls inside the 1$\sigma$ interval and the 1 $\sigma$ confidence interval is less than 50\% of the input prior range.}}
	\leftline{\footnotesize{\bm{$\circleddash$}: Part of parameters are constrained.}}
	
\end{table*}

Here, in order to see what kind of prior information is critical to resolve the above-mentioned two types of degeneracies, we run the $CS_{\rm 3bands}$ model by fixing different sets of parameters ($g$, $g\ \&\ P_{s}$, $\omega\ \&\ g_{0}$, and $f_{\rm cld}$) and evaluate the constraints on different aspects of the planet (bulk properties, gas mixing ratio, cloud vertical distribution, cloud optical properties and fraction, and surface spectral albedo). The parameters are fixed to fiducial values when they are assumed a priori, which represents the most optimistic situation.
Table 3 summarizes the results of the parameters that are usefully constrained within 1 $\sigma$ intervals when different sets of parameters are fixed. We find that a trustworthy estimation for atmospheric mixing ratios can be obtained by fixing $g\ \&\ P_{s}$, $\omega\ \&\ g_{0}$, or $f_{\rm cld}$. However, the known bulk properties ($g$ and $P_{s}$) offer no help in reducing the degeneracy that influences the overall spectra reflectance. This degeneracy can be reduced with the known cloud scattering properties or cloud fraction. Under these cases, the bulk properties, atmospheric mixing ratios, and the cloud vertical distribution can be well-constrained, although there are small biases of cloud optical properties and surface spectral albedo.
Additional observational information (e.g., polarization) could potentially be used to constrain cloud optical properties and surface spectral albedo. The varying cloud scattering properties of $\omega$ and $g_{0}$ and surface spectral albedo have unique responses in the polarization observation \citep{rossi2017}. 

\subsection{Cloud parameterization}
We adopt a simple parameterization to capture the cloud properties that produce the most significant effects on spectra \citep{feng2018}. Specifically, we allow for a single cloud deck with homogeneous distribution but varying cloud cover fractions in the atmosphere. In reality, clouds on Earth have an inhomogeneous distribution and variations in time \citep{cowan2009,fujii2013}.

\citet{mukherjee2021} found that the retrieved atmospheric and scattering properties strongly depend on the choice of cloud parameterization, and the necessity of increasing cloud complexity is highly sensitive to the relative position of the molecular, cloud, and Rayleigh scattering cross-section. We note that given the overall inability of simple cloud parameterization retrievals to constrain cloud and surface parameters (see also \citealp{lupu2016,nayak2017,feng2018}), increasing levels of cloud complexity may virtually add the degree of degeneracy between model parameters given limited wavelength-coverage and noise levels data.

Recent work has been undertaken to combine cloud microphysics models with 3D circulation models and to use this to predict emergent spectra \citep{lee2017,line2016,mendoncca2018}. It is imperative to conduct retrievals of simulated datasets based on more detailed, physically motivated, 3D cloudy atmosphere models and test whether the simple cloud parameterizations can represent the 3D cloudy models and accurately retrieve other atmospheric properties. 

\section{Summary}

In this paper, we explored the coupling effect brought by cloud and wavelength-dependent surface albedo on the performance of retrieval. The effects of multi-epoch observation, additional UV band data, and having estimations of cloud fraction on the improvement of parameter estimation have been explored. Our main conclusions are summarized below. 
\begin{enumerate}
	
\item When the cloud scattering properties ($\omega$, $g_{0}$) can be assumed a priori, our three photometric bands surface albedo parameterization could constrain atmospheric mixing ratios, surface albedo ($A_{r}$), and the key cloud parameters ($P_{b}$, $d_{p}$, $\tau$, $f_{\rm cld}$). With S/N = 20, different surface compositions (vegetation, ocean, barren world) can be distinguished. 

\item If we try to estimate the surface albedo using gray spectra while the true surface albedo is wavelength-dependent, the retrieval causes vastly biased results of atmospheric and cloud properties.

\item When the cloud scattering properties ($\omega $, $g_{0}$) cannot be assumed a priori, the inferred solution does not match the true value because of the degeneracy between the cloud properties ($\omega$, $g_{0}$, $f_{\rm cld}$) and the surface spectral albedo. The multi-epoch observations at the phases of $\frac{\pi}{2}$ and $\frac{\pi}{4}$ offer limited help in disentangling this degeneracy.

\item Even in the case of unknown cloud scattering and surface properties, the bulk and atmospheric properties (e.g., surface pressure, surface gravity, gas mixing ratios) can be reasonably inferred from the spectrum with a UV band (0.3 - 0.5 $\mu$m; R $\sim$  6) $+$ visible band (0.5 - 0.9 $\mu$m; R $\sim $ 140) combination. If short bandpass data is not available, then partial constraints on cloud coverage need to be inferred a priori with an uncertainty of 20\%.

\end{enumerate}

\acknowledgments{This work was supported by the National Natural Science Foundation of China (Grant No.11773045, 11973009, and 11933005) and the CAS Pioneer Hundred Talents Program. The authors also thank the anonymous referee for the instructive suggestions that substantially improved the quality of the paper.}

\bibliography{ref}
\bibliographystyle{aasjournal}

\appendix
\section{$C_{\rm \omega, g_{0}}S_{\rm 3bands}$ model: the impact from noise levels}

The initial choice of S/N = 20 is motivated by earlier studies \citep{lupu2016,nayak2017,feng2018,hu2019} establishing this to be the minimum S/N required for accurate retrievals of atmospheric properties. We also explore the effect of degrading S/N to 5 since we believe that the combinations of multi-epoch observations would push the limit of S/N requisite for accurate retrievals of atmospheric properties to a lower boundary \citep{damiano2020multi}.

Figure A.1 presents the comparison of the retrieved model parameters computed at S/N of 5 and 20 for both the $C_{\rm \omega, g_{0}}S_{\rm 3bands}$ and the $C_{\rm \omega, g_{0}}S_{\rm 3bands}\_2phases$ model. In the case of S/N = 5, the single-phase observation ($C_{\rm \omega, g_{0}}S_{\rm 3bands}$ model) present a posterior distribution that does not contain information, relative to the initially assumed prior (blue contours), which result in biases. However, we can still acquire the meaningful parameter constraints by combining multi-epoch observations (green contours). The constraints of the key parameters of bulk properties, atmospheric composition, and the cloud properties are consistent with input values to within 1$\sigma$. Although the distribution is more diffuse than the high S/N case (blue contours), the true value can be clearly found in the middle of the distribution. Adding a second observation at a different orbital phase ($\alpha=\frac{\pi}{4}$ in this study) at a relatively low S/N of 5 would drastically tighten the constraint of cloud and atmospheric properties. 

The effects of varying noise levels on each parameter are further demonstrated in Figure A.2, where the marginalized posterior distributions of various parameters with different S/N are summarized. As expected, the uncertainties on the model properties increase as the reflectance spectral noise gets larger. For most of the key parameters, lower data precision not only increases parameter uncertainties but also drives a “shift” in the parameter distributions. For example, lower precision observations drive the retrieval toward a clearer atmosphere with a lower atmospheric gas ratio and less abundant cloud while with relatively high surface albedo. This is because the large noises obscure the gas absorption. These subtle differences in the simulated spectra result in inaccuracies in retrieved values of atmospheric composition. There are also bimodal structures presented in the posterior distribution of gas mixing ratios (H$_{2}$O, O$_{3}$, O$_{2}$). The larger peak is around the true value, and the lower one represents the discrepancies brought by lower precision data. These underestimated values of gas mixing are compensated by a clearer atmosphere (lower cloud fraction). At S/N of 5, adding a second observation ($\alpha=\frac{\pi}{4}$) could effectively eliminate the bimodal structure of the posterior distribution of atmospheric composition.

In short, the S/N requirement for the $C_{\rm \omega, g_{0}}S_{\rm 3bands}$ model (single-phase observation) to give meaningful constraints on all model parameters is $\sim$ 15, which is consistent with the earlier spectral retrieval works \citep{lupu2016,feng2018}. However, interestingly, in the case of two-phases joint-retrieval, S/N $\sim$ 5 is enough to give meaningful parameter constraint. For instance, the H$_{2}$O abundance can be constrained to the 1$\sigma$ level, and the 1$\sigma$ constraint interval can be tightened to 16\% of the input prior ranges.
\begin{figure*}[!htb]
	\centering
	\includegraphics[height=14cm,width=18cm]{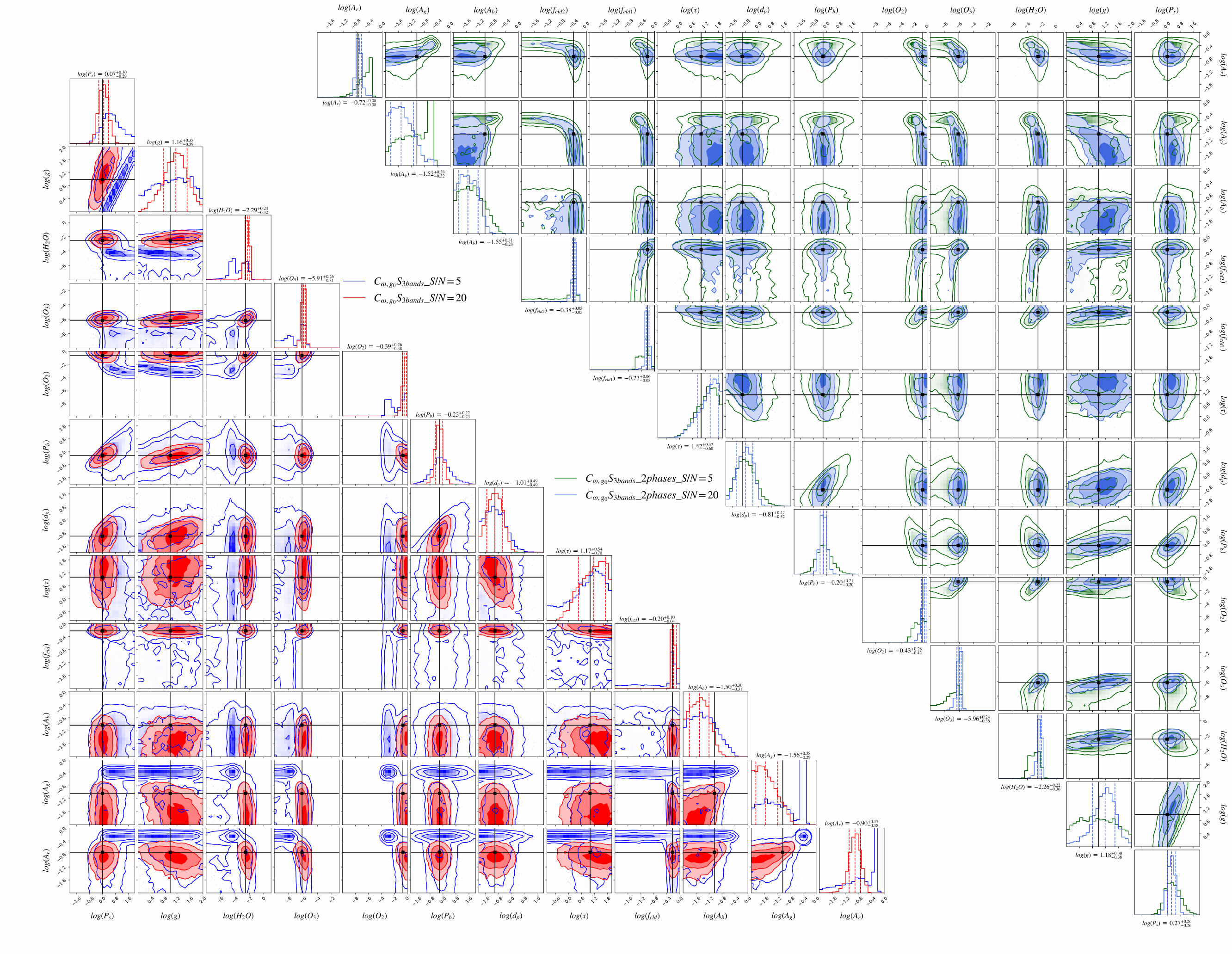}
	\caption{2D marginal posterior probability distributions for S/N = 5, 20.  Overplotted in black are the fiducial parameter values. Above the 1D marginalized posterior for each parameter, we list the median retrieved value with uncertainties of experiments with S/N= 20. Dashed lines (left to right) mark the 16\%, 50\%, and 84\% quantiles. Lower left: the $C_{\rm \omega, g_{0}}S_{\rm 3bands}$ model; Upper right: the $C_{\rm \omega, g_{0}}S_{\rm 3bands}\_2phases$ model.}
	\label{img14}
\end{figure*}

\begin{figure*}[!htb]
	\centering
	\includegraphics[height=22cm,width=18cm]{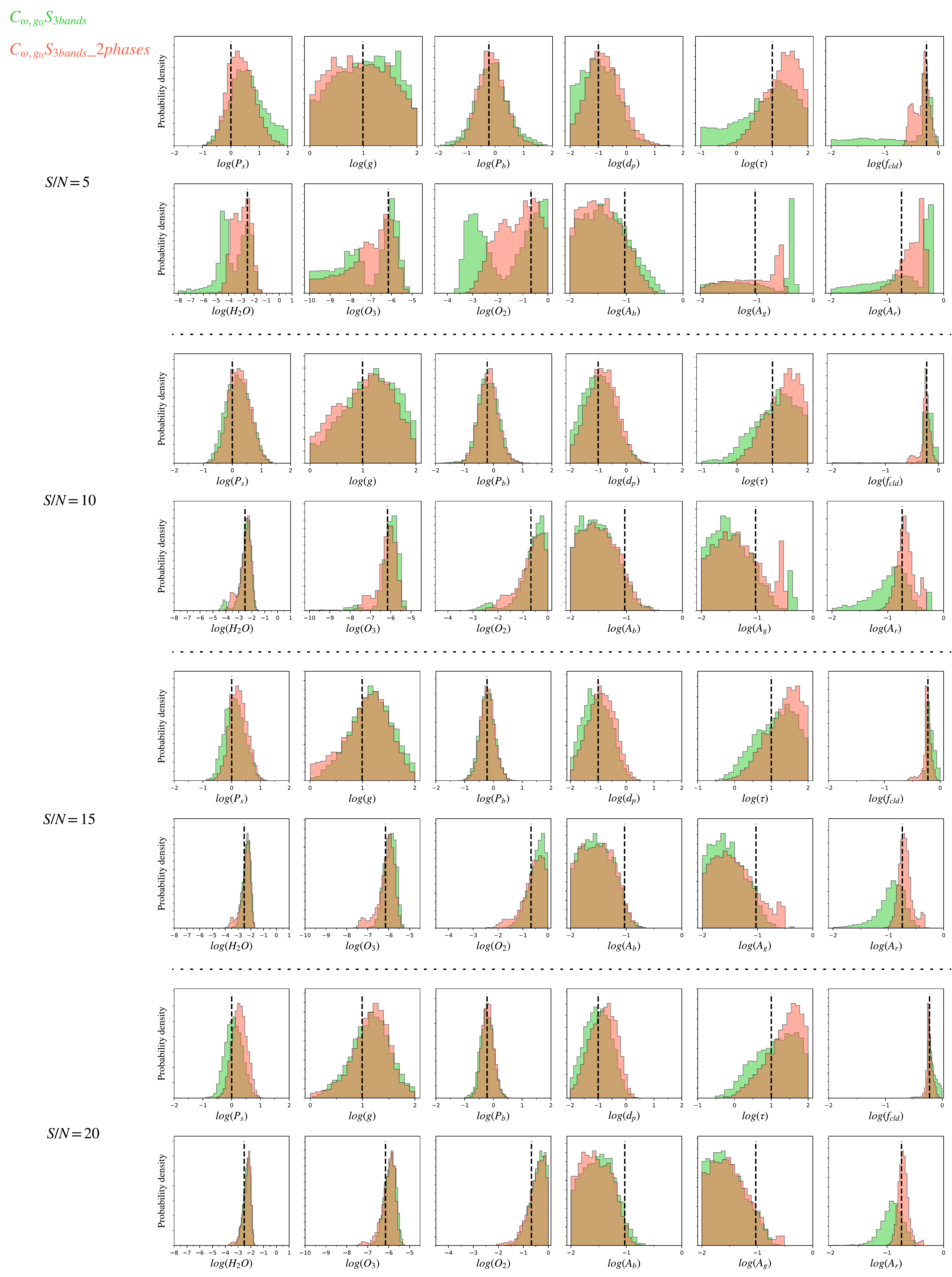}
	\caption{Retrieved parameters from simulated LUVOIR observations of Earth-like exoplanet. The predicted constraints from $C_{\rm \omega, g_{0}}S_{\rm 3bands}$ (green) and $C_{\rm \omega, g_{0}}S_{\rm 3bands}\_2phases$ (tomato) are shown by the marginalized posterior histograms. Each two rows correspond to different noise levels, ranging from 5 to 20. The true values are shown with black vertical lines.}
	\label{img15}
\end{figure*}

\end{document}